\documentclass[aps,prb,floatfix,twocolumn,superscriptaddress,showpacs,amsmath,amssymb]{revtex4}
\usepackage{graphicx}
\usepackage{epsfig} 
\allowdisplaybreaks

\newcommand{\be}{\begin{equation}}
\newcommand{\ee}{\end{equation}}
\newcommand{\bea}{\begin{eqnarray}}
\newcommand{\eea}{\end{eqnarray}}
\newcommand{\ba}{\begin{eqnarray*}}
\newcommand{\ea}{\end{eqnarray*}}
\newcommand{\dagga}{{\phantom{\dagger}}}
\newcommand{\bR}{\mathbf{R}}

\newcommand{\bq}{\mathbf{q}}
\newcommand{\bqp}{\mathbf{q'}}

\newcommand{\bRp}{\mathbf{R'}}

\newcommand{\dis}{\displaystyle}

\newcommand{\fract}[2]{\frac{\dis #1}{\dis #2}}

\newcommand{\eqn}[1]{(\ref{#1})}
\newcommand{\m}[1]{\mathcal{#1}} 
\newcommand{\eps}{\varepsilon}

\newcommand{\s}[2]{\sigma^{#1}_{#2}}
\newcommand{\aver}[1]{\langle {#1} \rangle}
\newcommand{\De}[1]{\Delta_{{#1}}(\bq)}
\newcommand{\DDe}[1]{\dot{\Delta}_{{#1}}(\bq)}

\begin{document}
\title{Linear Ramps of Interaction in the Fermionic Hubbard Model} 

\author{Matteo Sandri}
\affiliation{
International School for Advanced Studies (SISSA), and CRS Democritos, CNR-INFM, - Via Bonomea 265, I-34136 Trieste, Italy} 

\author{Marco Schir\'o} 
\affiliation{Princeton
  Center for Theoretical Science and Department of Physics, Joseph Henry
  Laboratories, Princeton University, Princeton, NJ
  08544} 

\author{Michele Fabrizio} 
\affiliation{
International School for Advanced Studies (SISSA), and CRS Democritos, CNR-INFM, - Via Bonomea 265, I-34136 Trieste, Italy} 
\affiliation{The Abdus Salam
  International Centre for Theoretical Physics (ICTP), P.O.Box 586,
  I-34014 Trieste, Italy}

\date{\today} 

\pacs{71.10.Fd, 05.30.Fk, 05.70.Ln}

\begin{abstract}
We study the out of equilibrium dynamics of the Fermionic 
Hubbard Model induced by a
linear ramp of the repulsive interaction $U$ from the metallic 
state through the Mott transition. 
To this extent we use a time dependent Gutzwiller variational method and 
complement this analysis with the inclusion of quantum fluctuations at the 
leading order, in the framework of a $Z_2$ slave spin theory.
We discuss the dynamics during the ramp and the issue of adiabaticity through  
the scaling of the excitation energy with the ramp duration $\tau$. 
In addition, we study the dynamics for times scales longer than the ramp time, 
when the system is again isolated and the total energy conserved. 
We establish the existence of a dynamical phase transition analogous to 
the one present in the sudden quench case and discuss its properties 
as a function of final interaction and ramp duration. 
Finally we discuss the role of quantum fluctuations on the mean field 
dynamics for both long ramps, where spin wave theory is sufficient, 
and for very short ramps, where a self consistent treatment of 
quantum fluctuations is required in order to obtain relaxation.

\end{abstract}
\maketitle

\section{Introduction}

Non equilibrium phenomena in closed quantum many body systems 
have recently become a very active field of research 
due to the experimental advances in trapping and cooling atomic 
quantum gases at extremely low temperatures.~\cite{Bloch_rmp08} 

Unlike standard solid state set-ups, experiments 
with ultra cold atoms feature an excellent degree of tunability as well as  
a very good thermal isolation from the environment, which  
make them perfect ground tests for studying time dependent phenomena and 
non equilibrium physics in strongly correlated 
quantum systems
~\cite{Greiner_nature_02_bis,Demler_doublons_prl10,Bloch_expansion_arXiv}.

In a typical cold atom experiment, microscopic parameters 
controlling the Hamiltonian of a quantum many body system, 
for instance the lattice depth or the interparticle interaction, 
are changed in time between different values 
following some given protocol~\cite{DeMarco_PRL11}. 
The dynamics during and after this time dependent transformation is recorded.  

From a theoretical perspective, if 
the rate of change is much faster than any typical time scale of the system, 
one can model such a process as a sudden change of parameters, 
a so called quantum quench~\cite{Calabrese_Cardy}. 
The interest on this class of non equilibrium phenomena has recently grown, 
triggering a novel debate on fundamental issues in quantum statistical 
mechanics, such as ergodicity and thermalization in closed quantum 
many body systems~\cite{SilvaRMP}.
Beside the general issue of thermalization and its 
relation to integrability~\cite{Olshanii_Rigol_nature,Biroli_Corinna_prl10} 
and  localization~\cite{Rossini_MBL,Carleo_MBL},  an intriguing question 
which has been recently addressed in a number of works concerns 
the ways strongly correlated system approach equilibrium, 
namely the short-to-intermediate time dynamics.  Here non trivial behaviours, 
featuring metastable prethermal 
states~\cite{Kehrein_prl08,Moeckel2009,Demler_pretherm11} 
trapping the dynamics for long time scales, are likely to emerge 
as a result of strong correlations. The intriguing possibility of 
sharp crossovers among different relaxation regimes, 
or even genuine dynamical transitions, has been firstly 
argued in a DMFT investigation of the fermionic Hubbard 
model~\cite{Werner_prl09} and then found in a number of mean field models, 
including interacting fermions~\cite{SchiroFabrizio_prl10}, 
bosons~\cite{SciollaBiroli_prl10}, spins~\cite{SciollaBiroli_long} and scalar 
fields~\cite{GambassiCalabrese_epl11}.

A rather different situation may arise if the time dependent protocol 
is performed in a finite time $\tau$, the simplest example being a 
linear-in-time increase of some control parameter, 
a so called  ramp. Here the hamiltonian of the system 
is explicitly time dependent and one may wonder about 
new issues concerning, 
for example, the degree of adiabaticity of the dynamics, 
namely to which extent an isolated system is able to follow a (slow) 
time dependent change of its Hamiltonian parameters without being 
excited~\cite{Polkovnikov_natphys}. Such a question has been around since the early days 
of quantum mechanics~\cite{BornFock}, an example being the 
Landau Zener process
~\cite{Landau_2levels,Zener_2levels,Majorana_2levels,Stueckelberg_2levels} 
where a two level system is driven through an avoided level crossing. 
In the context of quantum many body systems with a continuum of energy levels, 
this very basic idea lays the ground for the 
Landau's phenomenological description of 
Normal Fermi Liquids~\cite{Nozieres_book}.
More recently, the interest in the adiabatic dynamics of 
quantum many body systems has grown stimulated 
by a debate on quantum computation and mainly 
in connection with ramps across quantum critical points. 
In the small excitation energy limit, namely for slow ramps, 
the possibility of a universal behaviour has been discussed in a number 
of works~\cite{Polkovinkov_ramps,DeGrandi_ramps_short} as a generalization to 
isolated quantum systems of the classical dynamical behavior.

It is worth noticing at this point that understanding the 
degree of adiabaticity of a time dependent process in a 
quantum many body system is not only of theoretical interest 
but also of practical relevance for cold atoms applications. 
Indeed, one has to consider that real experiments are always performed 
at a finite rate which unavoidably induces heating into the system. 
Hence the challenge one has to face in order to use cold atoms 
to \emph{simulate} specific low temperature quantum phases 
is to minimize those heating effects. Recent works address 
this issue and look for the optimal ramping protocol which 
produces the minimal heating~\cite{Werner_ramps11,Chamon_ramps11}.
Other investigations on the slow quench dynamics in trapped cold gases
address the issue of equilibration of local and global quantities~\cite{Kollath_ramps11,Natu_ramps11}.

Finally, we note that while those questions mainly address the 
dynamics during the ramp, there are interesting issues as well 
that concern the evolution of the system once the ramp is over, 
namely for times $t>\tau$.  Here the system is again isolated, 
initialized with the excitation energy acquired during the ramp, 
and it is let evolve with its unitary dynamics. One can see that this set up 
is very similar to the quench case, with the ramp process affecting 
the initial condition of the dynamics. As we discussed, 
an interesting question in this case is to understand how 
the excitation energy due to the ramp affects the relaxation toward 
equilibrium and the possible existence of non trivial dynamical behaviours.

In this paper we address some of these questions in the context of the 
fermionic Hubbard model, which represents a paradigmatic example of strongly 
correlated system and it is of direct interest for cold atoms experiments. 
In particular,  by using the time dependent Gutzwiller approach we have 
recently developed~\cite{SchiroFabrizio_prl10,SchiroFabrizio_PRB11}, we will 
study linear ramps of the Hubbard interaction across the Mott transition. 
The paper is structured as follows. In section \ref{sect:sect2} 
we introduce the model and briefly review the literature on interaction ramps. 
In section \ref{sect:sect3} we introduce the time dependent Gutzwiller and 
discuss the results for the mean field dynamics. In section \ref{sect:sect4} 
we go beyond Gutzwiller using a slave spin formulation. Finally, 
section \ref{sect:sect5} is devoted to concluding remarks.

\section{Ramping the interaction in the Hubbard Model}\label{sect:sect2}

We consider the dynamics of the fermionic Hubbard model, whose Hamiltonian reads
\be\label{eqn:hubb}
\m{H}\left(t\right)= -\sum_{\sigma}\sum_{\langle \mathbf{R},\mathbf{R}'\rangle}\,t_{\mathbf{R}\mathbf{R}'}\,
 c^{\dagger}_{\mathbf{R}\sigma}\,c^\dagga_{\mathbf{R}'\sigma}+\frac{U\left(t\right)}{2}\,\sum_{\mathbf{R}}\,
\left(n_\mathbf{R}-1\right)^2\,
\ee
after a linear ramp of the interaction $U(t)$ between $U_i$ and 
$U_f=U_i+\Delta U$, namely we shall assume 
\bea
U(t)&=& U_i + \Delta U\,t/\tau\,\qquad 0<t<\tau \\
U(t)&=& U_f\,\qquad t\geq \tau \nonumber
\eea
We note that, experimentally, it turns to be easier to change in time 
the optical lattice depth, which controls the hopping strength $t_{\bR\bRp}$, 
rather than the local interaction. However, we can safely assume that the same effect can be modelled by tuning in time the local interaction, since the physics will only depend on the ratio between $U(t)$ and the bandwidth. In the following, 
we shall only focus on the half filled case and, for the sake of simplicity, 
consider a non interacting initial state ($U_i=0$), 
even though the extension to finite $U_i$ is straightforward.

The problem of linear ramps in a strongly correlated fermionic system has been addressed in a number of recent works. The crossover from adiabatic to sudden quench regimes and in particular the scaling of the excitation energy with the ramp time $\tau$ has been studied in the Falikov Kimball model by 
non equilibrium DMFT~\cite{Eckstein_ramp_FK}. For what concerns the 
Hubbard model,~(\ref{eqn:hubb}) the problem has been tackled 
in the perturbative small $U_f$ regime and arbitrary ramp-time 
using Keldysh perturbation theory~\cite{Kehrein_ramps}, and in the 
non-perturbative regime but short ramp times by non equilibrium DMFT 
in combination with CTQMC~\cite{Werner_ramps11}.  
Here we will make use of the mean field theory plus fluctuations we have 
developed for the sudden quench case to address the problem of ramps and we 
will compare with the results available whenever this is possible. 

Since the time dependent interaction $U(t)$ introduces a new time scale 
contrary to the sudden quench case, namely the rate $\tau$ at which the ramp 
is performed, one can ask oneself three separate questions: (i) 
what is the dynamics during the ramp, i.e. for times $t\le\tau$; 
(ii) what is the state the system is left once the ramp is terminated 
(excitation energy, degree of adiabaticity); and finally (iii) 
what is the non equilibrium dynamics for times larger than the ramp time, 
i.e. for $t>\tau$.

\section{Time Dependent Gutzwiller}\label{sect:sect3}
We start this section briefly reviewing the time dependent Gutzwiller approach 
we have recently developed\cite{SchiroFabrizio_prl10,SchiroFabrizio_PRB11} 
to describe real-time dynamics in correlated 
\emph{fermionic} systems (for a related approach to correlated bosons see 
Refs.~\onlinecite{SciollaBiroli_prl10,SciollaBiroli_long,Snoek_EPL11}).

Similar in spirit to the conventional ground state variational scheme, the 
idea of a time dependent  method is to give an ansatz for the wave function 
evolved at time $t$, $\vert\Psi(t)\rangle$,  in terms of a set of time 
dependent variational parameters whose dynamics is set by requiring the 
stationarity of the real time action functional
\be
\m{S}[\Psi^{\dagger},\Psi] = \int\,dt\,\langle\Psi(t)\vert\,i\partial_t-\m{H}\vert\Psi(t)\rangle.
\label{def:S-action}
\ee
In equilibrium, a variational wave function which is known to effectively 
describe the physics of 
strongly correlated fermions close to the Mott transition is the one originally proposed by Gutzwiller~\cite{Gutzwiller_1,Gutzwiller_2,Brinkman_Rice}. 
Its natural extension to the time dependent case reads
\bea
|\Psi(t)\rangle &=& 
\prod_\mathbf{R} \mathrm{e}^{-i\mathcal{S}_\bR(t)}\, \m{P}_\bR(t) \,
|\Phi(t)\rangle \nonumber \\
&\equiv& \mathcal{P}(t)\,|\Phi(t)\rangle,\label{eqn:ansatz}
\eea
where $|\Phi(t)\rangle$ are time-dependent variational wavefunctions for which 
Wick's theorem holds, 
hence Slater determinants or BCS wavefunctions, while   
$\m{P}_\bR(t)$ and $\m{S}_{\bR\alpha}(t)$ are hermitian operators that act on the 
Hilbert space at site $\mathbf{R}$ 
and depend on the variables $\lambda_{\bR\alpha}(t)$ and $\phi_{\bR\alpha}(t)$:    
\bea
\m{P}_\bR(t) &=& \sum_{\alpha}\,\lambda_{\bR\alpha}(t)\,\m{O}_{\bR\alpha},
\label{eqn:def_amplitude}\\
\m{S}_\bR(t)&=&\sum_{\alpha}\,\phi_{\bR\alpha}(t)\,\m{O}_{\bR\alpha},
\eea
where $\m{O}_{\bR\alpha}$ can be any local hermitian operator. It follows that 
the average value of $\m{O}_{\bR\alpha}$
\be
O_{\bR\alpha} = \langle \Psi(t)|\,\m{O}_{\bR\alpha}\,|\Psi(t)\rangle,\label{def:O_ialpha}
\ee
is a functional of all the variational parameters.
We shall assume that it is possible to invert (\ref{def:O_ialpha})
and express the parameters $\lambda_{\bR\alpha}$ as functionals of all the $O_{\bRp\beta}$, $\phi_{\bRp\beta}$ as well as of the parameters that define $|\Phi(t)\rangle$.

The exact evaluation of the action $\m{S}$ over the correlated wave 
function (\ref{eqn:ansatz}) is still an highly non trivial task which, 
in general, cannot be accomplished exactly. Rather one has to use 
approximation schemes or evaluate it numerically, using for example a suitable 
time dependent extension of the variational Monte Carlo algorithm as recently 
done in Ref.~\onlinecite{Carleo_MBL} for the bosonic Jastrow wave-function.

In this respect, the Gutzwiller approximation gives a prescription 
for such a calculation, which is exact in infinite coordination 
lattices,\cite{Gebhard,Michele_PRB07_dimer} although it is believed to 
provide reasonable results also when the coordination is finite. 
To this extent we impose that 
\bea
\langle \Phi(t)|\,\m{P}_\bR^2(t)\,|\Phi(t)\rangle &=& 1,\label{Gutzwiller-cond-1}\\
\langle \Phi(t)|\,\m{P}_\bR^2(t)\,\m{C}_{\bR\alpha}\,|\Phi(t)\rangle &=& 
\langle \Phi(t)|\,\m{C}_{\bR\alpha}\,|\Phi(t)\rangle,\label{Gutzwiller-cond-2}
\eea
where $\m{C}_{\bR\alpha}$ is any bilinear form of the single-fermion operators 
at site $\bR$, $c^\dagger_{\bR a}$ and $c^\dagga_{\bR a}$ with $a$ the spin/orbital index. 
Provided Eqs.~\eqn{Gutzwiller-cond-1} 
and \eqn{Gutzwiller-cond-2} hold one can show that, in the limit of infinite lattice coordination or equivalently within the Gutzwiller approximation, 
the average value of any local operator $\m{O}_{\bR\alpha}$ is given by 
\cite{Michele_PRB07_dimer}
\bea
O_{\bR\alpha} &=& \langle \Psi(t)|\,\m{O}_{\bR\alpha}\,|\Psi(t)\rangle \label{av-value-local}=\nonumber\\
&=& 
\langle \Phi(t)|\,\m{P}_\bR(t)\,\mathrm{e}^{i\m{S}_\bR(t)}\,\m{O}_{\bR\alpha}\,\mathrm{e}^{-
i\m{S}_\bR(t)}\,
\m{P}_\bR(t)\,|\Phi(t)\rangle,\nonumber 
\eea
which can be easily computed by the Wick's theorem. For what concerns operators coupling different sites
$\bR\not=\bRp$ such as the hopping term one can show that the average over the wave function reads
\bea
&&\langle \Psi(t)|\,c^\dagger_{\bR\,a}\,c^\dagga_{\bRp\,b}\,|\Psi(t)\rangle =\nonumber \\
&&=\sum_{c d}\, Q^*_{\bR,ac}\,Q^\dagga_{\bRp,bd}\, 
\langle \Phi(t)|\,c^\dagger_{\bR\,c}\,c^\dagga_{\bRp\,d}\,|\Phi(t)\rangle,\nonumber \label{def:cdaggeri-cdaggaj}
\eea
where the matrix elements $Q_{\bR,ab}$ are obtained by solving
\bea
&& \langle \Phi(t)|\,\m{P}_\bR(t)\,\mathrm{e}^{i\m{S}_\bR(t)}\,c^\dagger_{\bR\,a}\,
\mathrm{e}^{-i\m{S}_\bR(t)}\,
\m{P}_\bR(t)\,c^\dagga_{\bR\,c}\,|\Phi(t)\rangle \nonumber \\
&&~~= 
\sum_b\,Q^*_{\bR,ab}\,\langle \Phi(t)|\,c^\dagger_{\bR\,b}\,c^\dagga_{\bR\,c}\,|\Phi(t)\rangle.
\nonumber
\label{def-R}
\eea  
Finally for what concerns the time derivative one can show that this reads~\cite{SchiroFabrizio_PRB11}
\be
i\langle \Psi(t)|\partial_t\Psi(t)\rangle = \sum_{\bR\alpha}\,\dot{\phi}_{\bR\alpha}\,O_{\bR\alpha} 
+i\langle \Phi(t)|\partial_t\Phi(t)\rangle,\label{diff-t}\nonumber
\ee
so that, all in all, the real time action reads
\bea
\label{eqn:action}
\m{S}[\Psi^{\dagger},\Psi] 
&=& \int\, dt\, \bigg( \sum_{\bR\alpha}\, \dot{\phi}_{\bR\alpha}\,O_{\bR\alpha} 
- E\left[\phi_{\bR\alpha},O_{\bR\alpha},\Phi\right] \nonumber\\
&& \phantom{ \int\, dt\, \sum_{\bR\alpha}\,} +i \langle \Phi(t)|\partial_t\,\Phi(t)\rangle\bigg),
\eea  
where the energy functional is given by
\bea 
\label{eqn:gutzwiller_energy}
E\left[\phi_\bR,D_\bR,\Phi\right] &=& \frac{U(t)}{2}\sum_{\bR}\,D_{\bR} +\nonumber\\
&& 
+\sum_{\langle \bR\bRp\rangle}\,Q_\bR^\dagga\,Q_\bRp^*\, w_{\bR\,\bRp}(t) + H.c.\, ,
\eea
with
\be
w_{\bR\bRp}(t)=-t_{\bR\,\bRp}\,\sum_\sigma\,\langle 
\Phi(t)\vert\,c^{\dagger}_{\bR\sigma}c^\dagga_{\bRp\sigma}\,\vert\Phi(t) \rangle,\label{def:w}
\nonumber
\ee
and 
\be
D_\bR = \langle\Psi(t)\,\vert\,\left(n_{\bR}-1\right)^2
\vert\Psi(t)\rangle \nonumber
\ee

The saddle point of $\m{S}$ in Eq.~\eqn{eqn:action} with respect to $\phi_{\bR\alpha}$ 
and $O_{\bR\alpha}$ is readily obtained by imposing
\bea
\dot{\phi}_{\bR\alpha} &=& \fract{\partial E}{\partial O_{\bR\alpha}},\label{eqn:eom_1}\\ 
\dot{O}_{\bR\alpha} &=& -\fract{\partial E}{\partial \phi_{\bR\alpha}},\label{eqn:eom_2}
\eea
showing that these pairs of variables act like classical conjugate fields with Hamiltonian $E$. 
As far as $|\Phi(t)\rangle$ is concerned, since it is either a Slater determinant or a BCS wavefunction, 
the variation with respect to it leads to similar equations as 
in the time-dependent Hartree-Fock approximation,\cite{NegeleOrland_book} namely, in general, non-linear single particle Schr{\oe}dinger equations. 

For what concerns the specific problem at hand, namely the single 
band Hubbard model with a time dependent interaction $U(t)$, following 
Ref.\onlinecite{SchiroFabrizio_prl10} we pose
\bea \label{eqn:P_i}
\m{P}_\bR(t) &=& \sum_{n=0}^2\,\lambda_{\bR,n}(t)\,\m{P}_{\bR,n}\,,\\
\label{eqn:S_i}
\m{S}_\bR(t) &=& \sum_{n=0}^2\,\phi_{\bR,n}(t)\,\m{P}_{\bR,n}\,,
\eea
where $\m{P}_{\bR,n}$ is the projector at site $\bR$ onto 
configurations with $n=0,\dots,2$ electrons. Then assuming a time independent 
Slater determinant as well as an homogeneous and non magnetic wave 
function~\cite{SchiroFabrizio_PRB11} we can further simplify the dynamical 
equations. At half filling the classical mean-field dynamics for the  double occupancy $D(t)$ and its conjugate variable $\phi(t)$ reads
\bea\label{eqn:1D_dyn}
\dot{D} = \frac{\bar{\eps}}{2}\,\fract{\partial Z}{\partial \phi}\\
\dot{\phi} = \frac{U(t)}{2} - \frac{\bar{\eps}}{2}\,\frac{\partial Z}{\partial D}
\eea
where $\bar{\eps}=\frac{U_c}{8}$ is the kinetic energy of the Fermi Sea in 
units of the critical repulsion $U_c$ for the zero temperature equilibrium Mott transition, 
while $Z=\vert\,Q\vert^2$ is the time dependent quasiparticle weight which reads 
(at half-filling) $Z[D,\phi] = 8D\,\left(1-2D\right)\,\cos^2\phi$. The above dynamics derives from a classical hamiltonian which reads
\be
E[D,\phi]=\frac{U(t)}{2}D-\frac{\bar{\eps}}{2}\,Z[D,\phi]
\ee

In the following sections we are going to analyze 
this dynamics for different ramp durations and final values of the 
interaction $U_f$.
\begin{figure}[t]
\begin{center}
\epsfig{figure=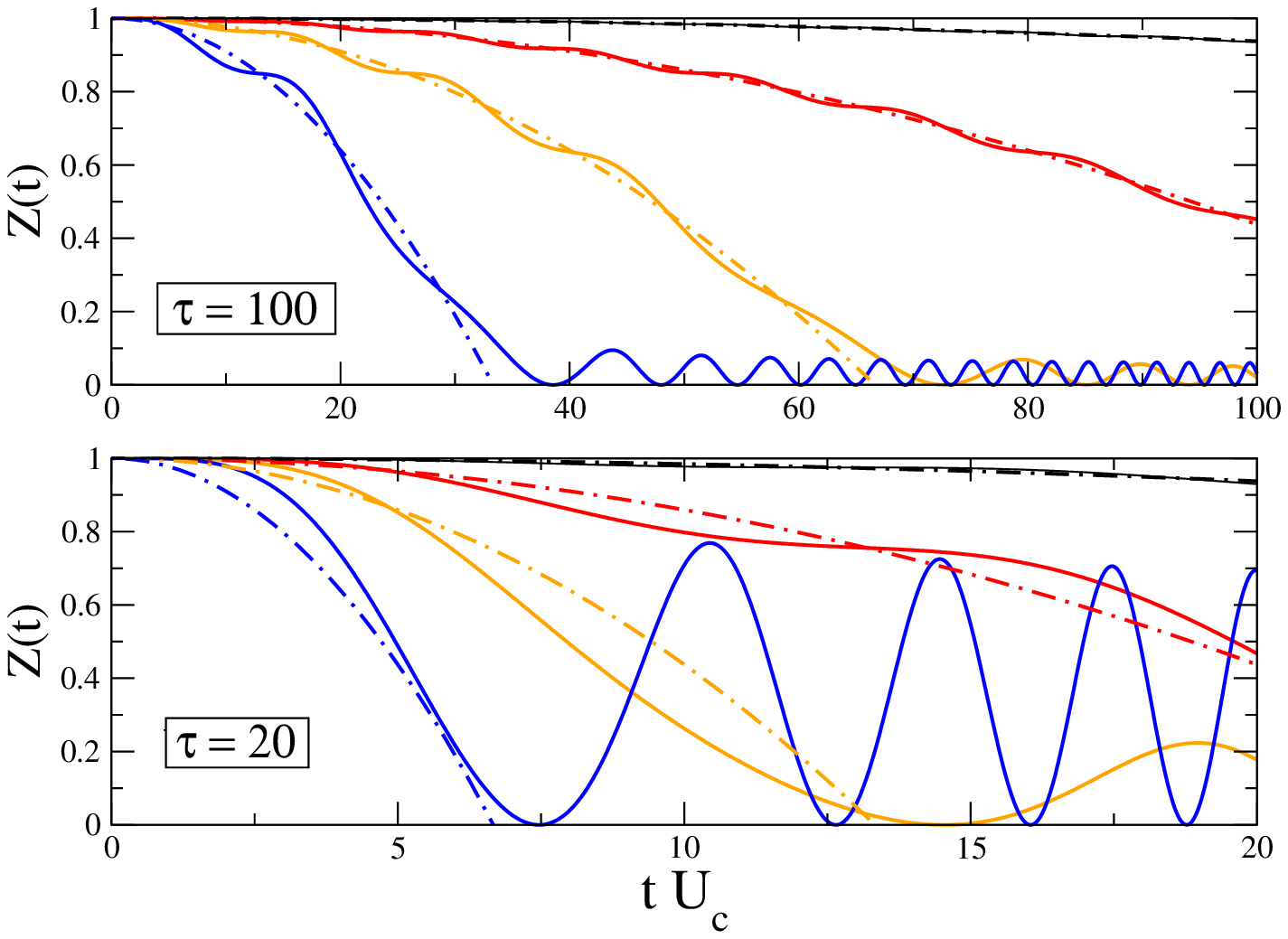,scale=0.31}
\caption{Gutzwiller mean field dynamics at half-filling for quasiparticle weight $Z(t)$ for quantum quenches from $u_i=0$ to $u_f=0.25,0.75,1.5,3.0$ (from top to bottom) for a ramp time $\tau=100$ (top panel) and $\tau=20$ (bottom panel). To comparison we plot the adiabatic dynamics $Z_{ad}(t)$ (see dashed lines), obtained assuming the system stays in its instantaneous variational ground state.}
\label{fig:fig1}
\end{center}
\end{figure}

\subsection{Dynamics during the ramp and degree of adiabaticity}


In Figure~\ref{fig:fig1} we plot the dynamics of the quasiparticle 
weight $Z(t)$ for different values of the final quench $u_f=U_f/U_c$ in units 
of $U_c$, the critical value for the equilibrium Mott transition (see after Eq.~(\ref{eqn:1D_dyn})) that will be our unit of energy hereafter,   at two different fixed ramp times, $\tau=100$ (top panel) 
and $\tau=20$ (bottom panel).  In the same figure we plot, for the sake of 
comparison, 
the adiabatic dynamics
obtained assuming the system stays in its instantaneous ground state, 
namely that
$$
Z_{ad}(t) = 1-u^2(t)\,.
$$
A quick look to this figure reveals that, as one could expect, 
the degree of adiabaticity depends strongly on the duration of the ramp 
$\tau$ and on the final value of the interaction $u_f$. 
In order to be more quantitative on this issue it is useful to introduce a 
measure of the adiabaticity of the process. A possible criterion amounts to 
calculate the excitation energy which is left into the system once the ramp is 
completed. This quantity is defined as
\be
\Delta E_{exc}\left(\tau,u_f\right) = E\left(\tau,u_f\right) - 
E_{gs}(u_f(\tau))\,,
\ee
where $E(t,u(t))=\langle H(t)\rangle$ is the time dependent expectation 
value of the Hamiltonian, while $E_{gs}(u_f)$ is the ground state energy 
at the final value of the interaction $u_f$. Based on very general 
grounds one expects that if the system 
behaves adiabatically then the excitation energy $\Delta E_{exct}$ should go 
to zero as the ramp duration diverges. Since one expects the process to 
be more and more adiabatic as $\tau$ increases, the expectation for 
$\Delta E_{exc}$ is to show a monotonic decreasing behaviour 
as a function of the ramp time $\tau$.

\begin{figure}[h]
\begin{center}
\epsfig{figure=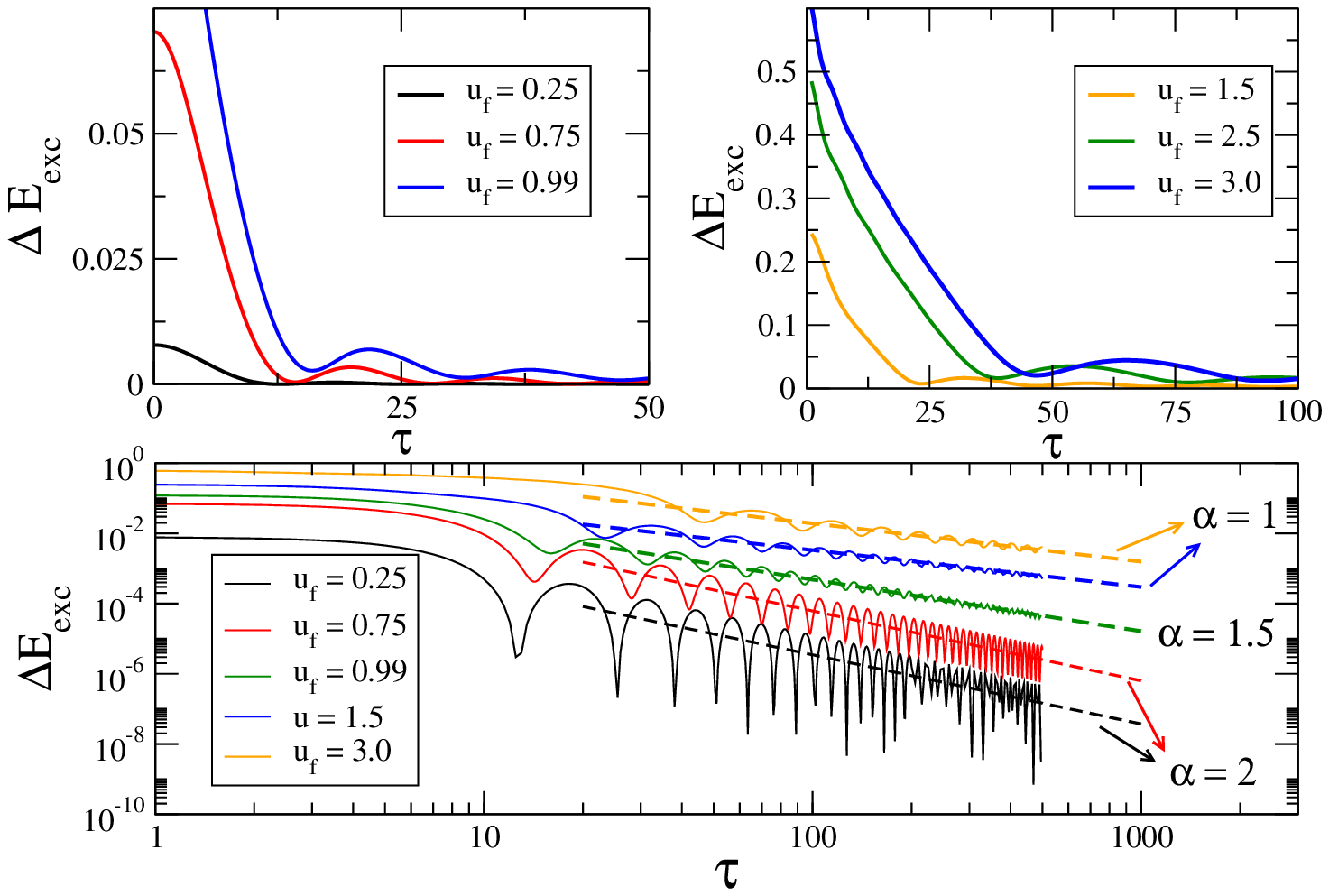,scale=0.32}
\caption{Excitation energy $\Delta\,E(\tau)$ as a function of the ramp time for quenches starting from the metallic phase ($u_i=0$) and ending into the metallic (left panel) or insulating (right panel) phase. We see in the former case a fast transient to zero occurs, with some residual oscillations which die out as $\tau$ increases. As opposite for quenches which crosses the Mott transition the transient seems much more longer and sensitive to the final value of $u_f$, namely stronger quenches seems to require longer ramps to achieve a fixed amount of excitation energy. }
\label{fig:fig2}
\end{center}
\end{figure}

In Figure~\ref{fig:fig2} (top panels)  we plot the excitation energy 
as a function of $\tau$ for quenches from the non interacting case $u_i=0$ 
to different values of $u_f$. We notice the
excitation energy does indeed decreases toward zero with $\tau$, 
although with some small oscillations, thus confirming that the time 
dependent Gutzwiller approximation is able to capture the crossover from the 
sudden quench to the adiabatic regime. 

It is particularly interesting to study the regime of very long ramp times $\tau\rightarrow\infty$, 
where one expects universal behaviour to emerge as a function of the ramp speed. This universality translates into power-laws and scaling relations for the relevant physical observables
which have been recently attracting a lot of attention in the literature, starting with the seminal work by Kibble and Zurek on classical phase transitions and its generalization to the quantum case~\cite{Dziarmaga_PRL05,Polkovinkov_ramps,Zurek_prl05}. More recently the issue of universality in the Kibble-Zurek problem has attracted a renewed interest and first steps toward a scaling theory   have been performed~\cite{Koludrubetz_PRL12,Chandran_arxiv2012}. Here we focus on the scaling of the excitation energy $\Delta E_{exc}$ which is very sensitive to the nature of the elementary excitations in the systems~\cite{Polkovnikov_natphys}. 
This question, in the context of the correlated fermionic systems, has been addressed in the Falikov-Kimball model using DMFT~\cite{Eckstein_ramp_FK} and in the fermionic Hubbard model, that is of interest here, mainly using pertubation theory~\cite{Kehrein_ramps,Eckstein_ramp_FK}. 

We perform such a scaling analysis (see bottom panel of figure ~\ref{fig:fig2}) and find that to a very good extent the behaviour of $\Delta E_{exc}$ is consistent with a power law, possibly with a pre-factor that depends on the interaction $u_f$ and displays in general an extra oscillating behaviour in $\tau$
\be\label{eqn:deltaE}
\Delta E_{exc}(\tau)=\frac{\gamma(\tau,u_f)}{\tau^{\alpha}} 
\ee

At small values of the final interaction $u_f$ we find 
$\Delta E_{exc}\sim \tau^{-2}$. We notice that, in this small quench regime, oscillations are more pronounced (and result into the noisy scaling of figure~\ref{fig:fig2}), nevertheless 
the power law scaling with $\alpha=2$ works very well for the envelope of 
local maxima. This scaling appears to be consistent with perturbative results~\cite{Kehrein_ramps,Eckstein_ramp_FK}
 and with linear response arguments~\cite{Polkovnikov_natphys}. We notice that for the Falikov-Kimball model the DMFT analysis gives a different exponent, $\alpha=1$, for ramps ending in the metallic phase, but this result has been understood as a consequence of the Non-Fermi Liquid ground state of that model~\cite{Eckstein_ramp_FK}. 
\begin{figure}[h]
\begin{center}
\epsfig{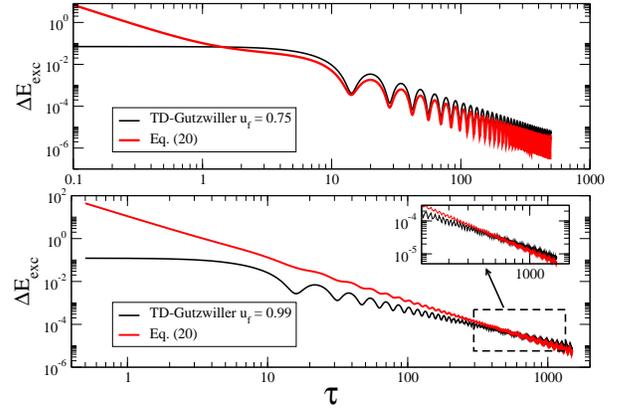}
\caption{Excitation energy $\Delta\,E(\tau)$ as a function of the ramp time for quenches starting ($u_i=0.0$) and ending ($u_f=0.75$ top panel, $u_f=0.99$ bottom panel) in the metallic phase We compare the Gutzwiller results with the scaling Eq.~(\ref{eqn:deltaE}-\ref{eqn:scaling_metal}) obtained from the adiabatic classical dynamics. The agreement for $u_f<1$ is excellent but worsten upon approaching the critical point. }
\label{fig:fig2bis}
\end{center}
\end{figure}
Within our time dependent Gutzwiller approximation we find that
 the ``Fermi Liquid scaling'' works up to rather large values 
of the interaction but appears to break down close to the Mott 
transition, $u_f\alt 1$, where the exponent crosses 
over to $\alpha\simeq1.5$. 
Finally, for ramps ending deep inside 
the Mott phase, we find very small oscillation in the long time behaviour 
of $\Delta E_{exc}$ and power law scaling suggests an exponent $\alpha=1$. 
In order to get more insights into the behaviour of the excitation energy $\Delta E_{exc}$ for large $\tau$ it is useful to step back for a moment to the Gutzwiller semi-classical dynamics given by equations~(\ref{eqn:1D_dyn}). 
In the limit of very slow ramps, $\tau\rightarrow\infty$, one can analyze the deviations from adiabaticity using techniques borrowed from classical mechanics. This is described in great detail in a recent work by Bapst and Semerjain that addresses the ramp dynamics in a fully connected $p-$spin model with a transverse field~\cite{Guilhem_ramps}. For ramps ending in the metallic phase, $u_f<1$, one can expand the classical hamiltonian around its instantaneous minimum~\cite{Guilhem_privatecomm} (see appendix~\ref{sect:appA}), $D_*(t)=(1-u(t))/4$, up to a quadratic order with frequency $\omega(t)\sim\sqrt{1-u(t)^2}$ and obtain for the excitation energy the result~(\ref{eqn:deltaE}) with $\alpha=2$ and $\gamma$ 
\bea\label{eqn:scaling_metal}
\gamma(\tau,u_f)= \frac{u_f^2\,\sqrt{1-u_f^2}}{4}\sin^2\,\omega(u_f)\tau+\nonumber\\+
\frac{1-u_f^2}{4}\left(\frac{u_f}{1-u_f^2}-\frac{u_f}{(1-u_f^2)^{1/4}}\cos\,\omega(u_f)\tau\right)^2
\eea
with $\omega(u_f)=\frac{1}{4}\left(\arcsin(u_f)/u_f+\sqrt{1-u_f^2}\right)$. In figure~\ref{fig:fig2bis} we compare this expression with the numerics and find an excellent agreement, in particular we notice the frequency of the oscillations is correctly captured by $\omega(u_f)$. We also notice that upon approaching the critical point $u_f\rightarrow 1$ the agreement deteriorates. Indeed for ramps ending in the insulating phase, i.e. $u_f>1$, the situation is more tricky as the frequency of oscillations $\omega(t)$ vanishes during the ramp at $t=t_{\star}=\tau/u_f$ and one cannot extend the above analysis to the regime $t_{\star}<t<\tau$. Still one can proceed by mapping the classical dynamics onto a suitable limit of the Painlev\'e equation and using the well known results on its asymptotic. This has been discussed in Refs~\cite{Itin_ramps_LMG,Guilhem_ramps} for the fully connected Ising model in a transverse field, which is relevant for the Hubbard model within the Gutzwiller approximation~\cite{SchiroFabrizio_PRB11}, where a power law $\alpha=1$ has been found for ramps across the critical point, in agreement with our numerical results.  In light of this analysis an interesting question, that we leave open for future investigations, is to understand whether a different power law exponent may arise for ramps ending right at the critical point (as our numerics would suggest) or if the quadratic scaling~(\ref{eqn:scaling_metal}) expected in the metallic phase eventually sets in on a sufficiently longer time scale.  

We finally conclude this section by briefly discussing whether the above findings can be put into the framework of the Kibble-Zurek scaling theory~\cite{Dziarmaga_PRL05,Polkovinkov_ramps,Zurek_prl05}. For a ramp from the ordered to the disordered phase across a critical point scaling arguments would predict for the excitation energy a power-law decay~\cite{DeGrandi_prb09,Pollmann_rampsPRBR2010} $\Delta E_{exc}\sim1/\tau^{d\nu/z\nu+1}$. Indeed, by using the mean field exponents $\nu=1/2,z=1$ for the Ising critical point and setting $d$ to the upper-critical dimension $d=3$ for a quantum Ising model we get $\Delta E_{exc}\sim 1/\tau$, namely $\alpha=1$, which matches our results. While this observation may suggest a positive answer to this question we notice that the validity of such a scaling theory for fully connected models (or finite-connectivity models treated within mean-field as it is the case here) is not obvious a priori (in particular the identification of $d$ with the upper critical dimension is generally dangerous when dealing with scaling) and it has been not fully addressed in the literature to the best of our knowledge. For this reason and since this is not the main focus of the present paper we refrain from conclusive statements on this issue and leave this question for future investigations.

\subsection{Dynamics after the ramp}

We now turn our attention on the dynamics \emph{after} the ramp is completed, 
namely for $t>\tau$. Here the system is isolated, i.e. the energy is 
conserved, and the evolution starts from the state the system is left 
once the ramp is over. This set-up represents therefore the natural 
generalization of the sudden quench case (which is indeed recovered in the 
limit $\tau\rightarrow0$): once the ramp is completed, the system has some 
excitation energy above its ground state and one is interested in the 
relaxation dynamics for longer time scales.

Interestingly enough this issue has been only partially addressed in the 
literature, which mostly focused on the dynamics during the ramp,  
but it looks particularly intriguing in light of the results obtained on the 
sudden quench case. As we mentioned in the Introduction, 
a dynamical transition characterized by a fast relaxation has been found, 
quite generically, in mean field models for bosons and 
spins~\cite{SciollaBiroli_prl10,SciollaBiroli_long} and in the fermionic case, 
too, both at the variational level~\cite{SchiroFabrizio_prl10} and within 
DMFT~\cite{Werner_prl09}.

A natural question we would like to address here is therefore what 
is the effect of the finite ramp duration on the mean field dynamical 
transition found in the sudden quench case. A recent investigation using non 
equilibrium DMFT with the CTQMC impurity solver~\cite{Werner_ramps11} 
addressed this same issue for very small ramps and found 
signatures of a sharp crossover in the dynamics, much similar to what 
found in the sudden quench limit. While this result seems to suggest that 
a dynamical transition survives also for small finite $\tau$, 
it is difficult from numerical data, which are limited to short times, 
to conclude what happens for a generic speed ramp, and eventually in the 
adiabatic limit $\tau\rightarrow\infty$.  Here we will address again this 
point using mean field theory and study the fate of the dynamical transition 
after a ramp of arbitrary speed. 

As we mentioned earlier, the classical dynamics~(\ref{eqn:1D_dyn}) 
for $t>\tau$ admits an integral of motion which is the total energy, 
\be
E(t) = u_f\,D(t)-\frac{1}{8}\,Z(t)\equiv E_R(u_f,\tau)\,,\qquad t>\tau\,,
\ee
hence we can use it to reduce the problem to a one dimensional dynamics, 
much in the same way we did for the quench case. A simple calculation gives
\be
\dot{D}=\sqrt{\Gamma(D)}\,,
\ee
with the effective potential $\Gamma(D)$ given by
\be\label{eqn:eff_pot}
\Gamma(D)=\left(u_f\,D-E_R\right)\Big(E_R-u_f\,D+2\,D\left(1/2-D\right)\Big)
\ee
The energy $E_R(u_f,\tau)$ after the ramp depends on 
the initial ($u_i$) and final ($u_f$) values of the interaction 
and from the ramp time $\tau$.  In the general case, 
its value has to be determined from the solution of the dynamics for 
$t<\tau$, but it reduces in the sudden quench limit ($\tau\rightarrow0$) 
to the value
$$
E_R(u_f,0^+)= \frac{u_f}{4}-\frac{1}{8}\,,
$$
while for an infinitely slow ramp $\tau\rightarrow\infty$ it approaches the 
ground state energy at the final value of the interaction, namely
$$
E_R(u_f,\tau\rightarrow\infty)=-\frac{1}{8}\left(1-u_f\right)^2\,\qquad u_f<1\,,
$$
and zero in the Mott insulator phase $u_f>1$.

\begin{figure}[t]
\begin{center}
\epsfig{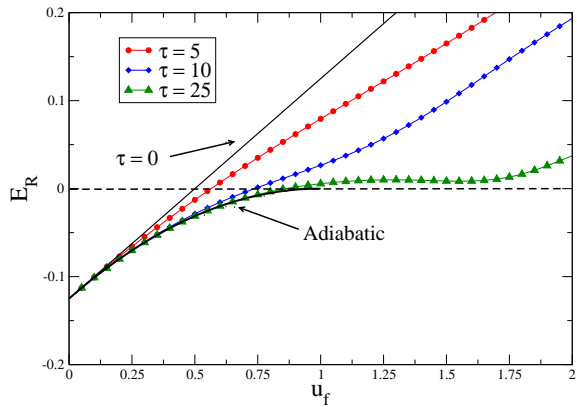}
\caption{Average energy after the ramp as a function of the final interaction quench $u_f$ and for different values of the ramp time $\tau$. We see that upon increasing $\tau$ the energy crosses over from the sudden quench limit to the adiabatic instantaneous ground state energy $E_{gs}(u_f)$.}
\label{fig:fig3}
\end{center}
\end{figure}

In Figure~\ref{fig:fig3} we plot the behaviour of $E_R(u_f,\tau)$ at different 
values of $\tau$.
 The effective potential has three roots which read $D_{\star} = E_R/u_f$ and
\bea\label{eqn:inversion}
D_{\pm} = \frac{1-u_f\pm\sqrt{\left(u_f-1\right)^2+8\,E_R}}{4}
\eea
\begin{figure}[t]
\begin{center}
\epsfig{figure=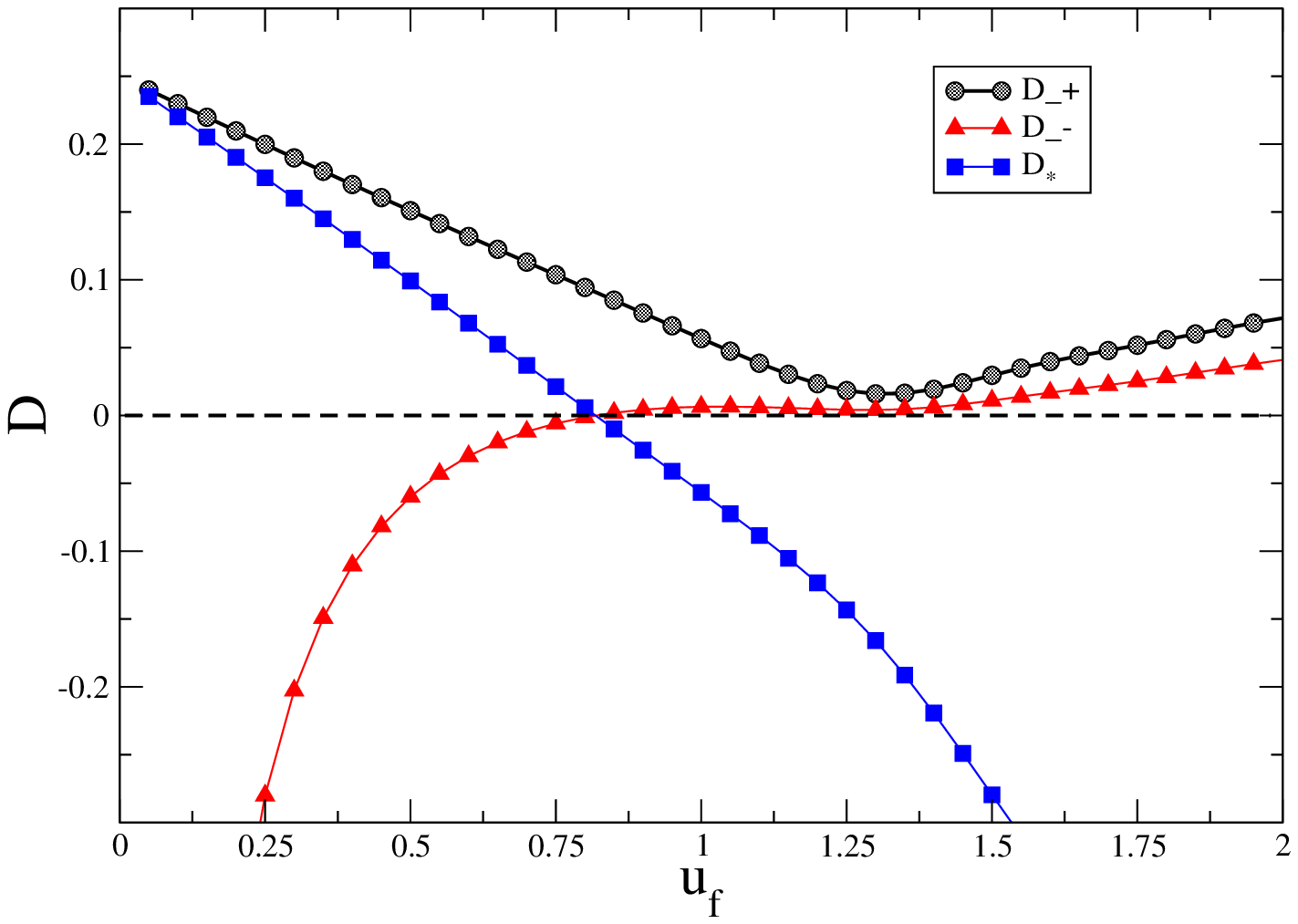,scale=0.31}
\caption{Behaviour of the inversion points $D_{\pm}, D_{\star}$ as a function of the final interaction $u_f$, for a ramp of the interaction of duration $\tau=20$ and starting from $u_i=0$. We notice the crossing of roots, occurring at $u_f^c(\tau)$ which signals the onset of a relaxation dynamics.}
\label{fig:fig4}
\end{center}
\end{figure}
We immediately see that, much as in the sudden quench case, for a given ramp 
time $\tau$ at which the condition $E_R(u_f,\tau)=0$ is fulfilled, 
two of the above roots merge and the dynamics shows an 
exponentially fast relaxation. The only non vanishing root reads
\be
D_{-} = \frac{1-u^c_f(\tau)}{2}
\ee
where $u_f^c(\tau)$ is the value of the final interaction at which 
$E_R(u_f,\tau)=0$.
Using the inversion points (\ref{eqn:inversion}), 
we can easily characterize the dynamics after the ramp, 
for different values of the final interaction $u_f$, 
in terms of period and amplitude of oscillations in the same way we 
did for the sudden quench case~\cite{SchiroFabrizio_prl10}. 
In Figure~\ref{fig:fig5} we plot the dynamics of quasiparticle weight $Z(t)$ 
after a ramp of $\tau=20$ for different values of the final interaction. 
We still can distinguish two regimes of slow and fast oscillations 
with some period $\m{T}(u_f,\tau)$, which turns out to diverge at the 
transition $u_f^c(\tau)$. Such a diverging time scale is associated to 
a change in the behaviour of the effective potential $\Gamma(D)$, with two 
inversion points going degenerate at $u_f^c(\tau)$. As a result, the 
divergence appears to be still logarithmic 
$\m{T}\sim \log\vert u_f-u_{f}^c(\tau)\vert$.
\begin{figure}[t]
\begin{center}
\epsfig{figure=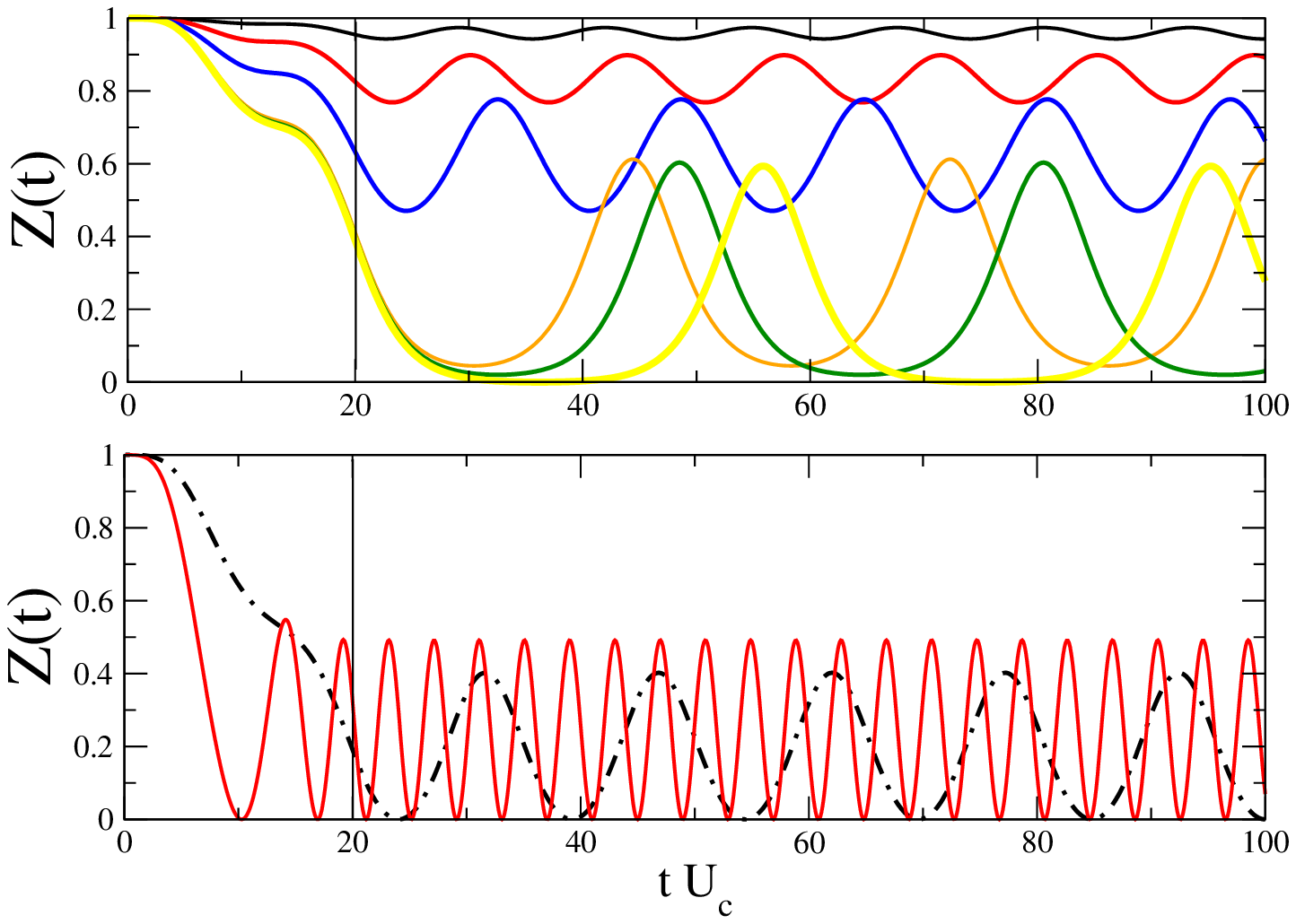,scale=0.31}
\caption{Gutzwiller mean field dynamics at half-filling for quasiparticle weight $Z(t)$ for a ramp of duration $\tau=20$ from $u_i=0$ to $u_f=0.2,0.4,0.6,0.8,0.81,0.82$ (top panel, from top to bottom) and from $u_i=0$ to $u_f=1.0$ (dashed) or $2.0$ (full red line). The critical value of the interaction quench $u_f^c(\tau=20)\simeq 0.83$. }
\label{fig:fig5}
\end{center}
\end{figure}
For ramps ending right at $u_f^c(\tau)$ the dynamics approaches exponentially fast the steady state value $Z=0$. The exact expression for $Z(t)$ 
can be worked out in this case, but does not look particularly illuminating . 
The scaling at long times reads
\be
Z(t\gg\tau)\sim \exp\Big(-t/t_{rel}\Big)
\ee
with $t_{rel}=\sqrt{2\,u_f^c(\tau)}$. We therefore get an exponential 
scaling at long times, as for the sudden quench case, with a time scale 
$t_{rel}$ that accounts for the finite duration of the ramp.
It is interesting to discuss the dependence of the critical interaction 
strength $u_f^c$ from the ramp duration $\tau$, which could shed some 
light on the origin of this putative dynamical critical point, which is 
still under debate. In addition to that, as we noticed earlier, 
this quantity (together with the lattice bandwidth) sets the time scale 
for the relaxation $t_{rel}$, therefore by tuning properly $\tau$ one can 
arrange protocols where relaxation is faster. 
This issue was addressed in Ref.~\onlinecite{Werner_ramps11}, 
although only for short ramps $\tau\simeq 1$, where the authors also 
discussed the dependence of $u_f^c(\tau)$ upon the ramp protocol.

\begin{figure}[t]
\begin{center}
\epsfig{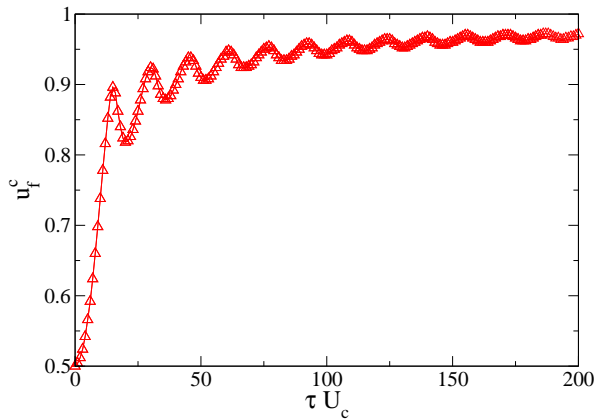}
\caption{Mean field dynamical critical point $u_f^c$ as a function of the ramp duration $\tau$. We see that for small $\tau$ we recover the sudden quench result $u_f^c=1/2$ while for longer ramps $\tau\rightarrow\infty$ $u_f^c$ approaches the equilibrium Mott critical point $u_f=1$.}
\label{fig:fig6}
\end{center}
\end{figure}

In Figure~\ref{fig:fig6} we plot the behaviour of $u_f^c$ 
as a function of $\tau$ for a linear ramp starting at $u_i=0$. 
We see that this quantity approaches, for $\tau\rightarrow0$, the sudden 
quench value $u_f^c(0)=1/2$.  From the behaviour of $E_{R}(u_f,\tau)$ 
in Figure~\ref{fig:fig3}  we observe that in the opposite limit of a 
very long ramp the system is closer and closer to the adiabatic ground state. 
As a result, the condition $E_R(u_f,\tau)=0$ suggests that as $\tau$ 
increases the mean field critical point $u_f^c$ smoothly approaches the 
equilibrium zero temperature Mott transition, 
namely $u_f^c(\tau\rightarrow\infty)=1$. 
This is indeed the case, namely $u_{f}^c$ interpolates between the 
sudden quench value at small $\tau$ and the Mott critical point for 
long ramps. We also note the presence of small oscillations in its $\tau$ 
dependence, which are likely an artefact of the Gutzwiller mean field 
dynamics. DMFT data would be required in order to check this point further.

The asymptotic behaviour for long ramps, namely for small excitation 
energies, looks also very intriguing and deserves further investigations. 
From one side one could have expected this result since the larger is 
$\tau$ the less the system is excited at the time the ramp stops. 
Hence it is reasonable to expect that some kind of \emph{criticality} 
or sharp crossover between weak and strong coupling should be visible close 
to the Mott quantum critical point. On the other hand, 
one has also to bear in mind that the less the system is excited above 
its ground state at the time the ramp ends, the less sharp the signature of 
the \emph{dynamical} critical point will look. Indeed, as we are going 
to see and in agreement of what observed by Keldysh perturbation theory,
~\cite{Kehrein_ramps} the 
metastable prethermal states which block the dynamics at small and large 
quenches become lower and lower in energy as the ramp time increases.


The last issue we would like to discuss here is the dependence 
from the ramp time $\tau$ of long time averages, that we define as
\be\label{eqn:long_time_aver}
\bar{O}=\mbox{lim}_{t\rightarrow\infty}\,\frac{1}{t}\,\int_{\tau}^t\,dt'\,O\left(t'\right)\,.
\ee 
Since the motion for $t>\tau$ is periodic, although as we have seen 
the initial condition at $t=\tau$ is not an inversion point of the dynamics, 
one can still express those long time averages as an integral over a period 
of oscillation. This allows to obtain closed expressions for the 
double occupation average $\bar{D}\left(u_f,\tau\right)$ and, through the 
conservation of energy, for the quasiparticle weight $\bar{Z}(u_f,\tau)$. 
It is then easy to see that both quantities vanish at the dynamical critical 
point $u_f^c(\tau)$ with the same logarithmic 
divergence found in the sudden quench case,
\be
\bar{D}(u_f,\tau)\sim 1/\log\vert u_f-u_f^c(\tau)\vert\,.
\ee
In other words, as for the period of oscillations $\m{T}(u_f,\tau)$, the only effect of the finite ramp duration is to shift the critical point to $u_f^c(\tau)$, without changing the critical behaviour at the transition.

In addition to the behaviour close to $u_f^c(\tau)$, also the 
results at small and large values of $u_f$ are interesting. Indeed, in the 
sudden quench case we have shown~\cite{SchiroFabrizio_PRB11} that the 
long time average of mean field dynamics exactly reproduces the metastable 
plateaux blocking the dynamics, which can be evaluated using  perturbation 
theory. This is consistent with the idea that mean field dynamics is able 
to capture the short-to-intermediate dynamics, and the trapping occurring on 
those time scales, but not the final escape toward equilibrium. 
\begin{figure}[t]
\begin{center}
\epsfig{figure=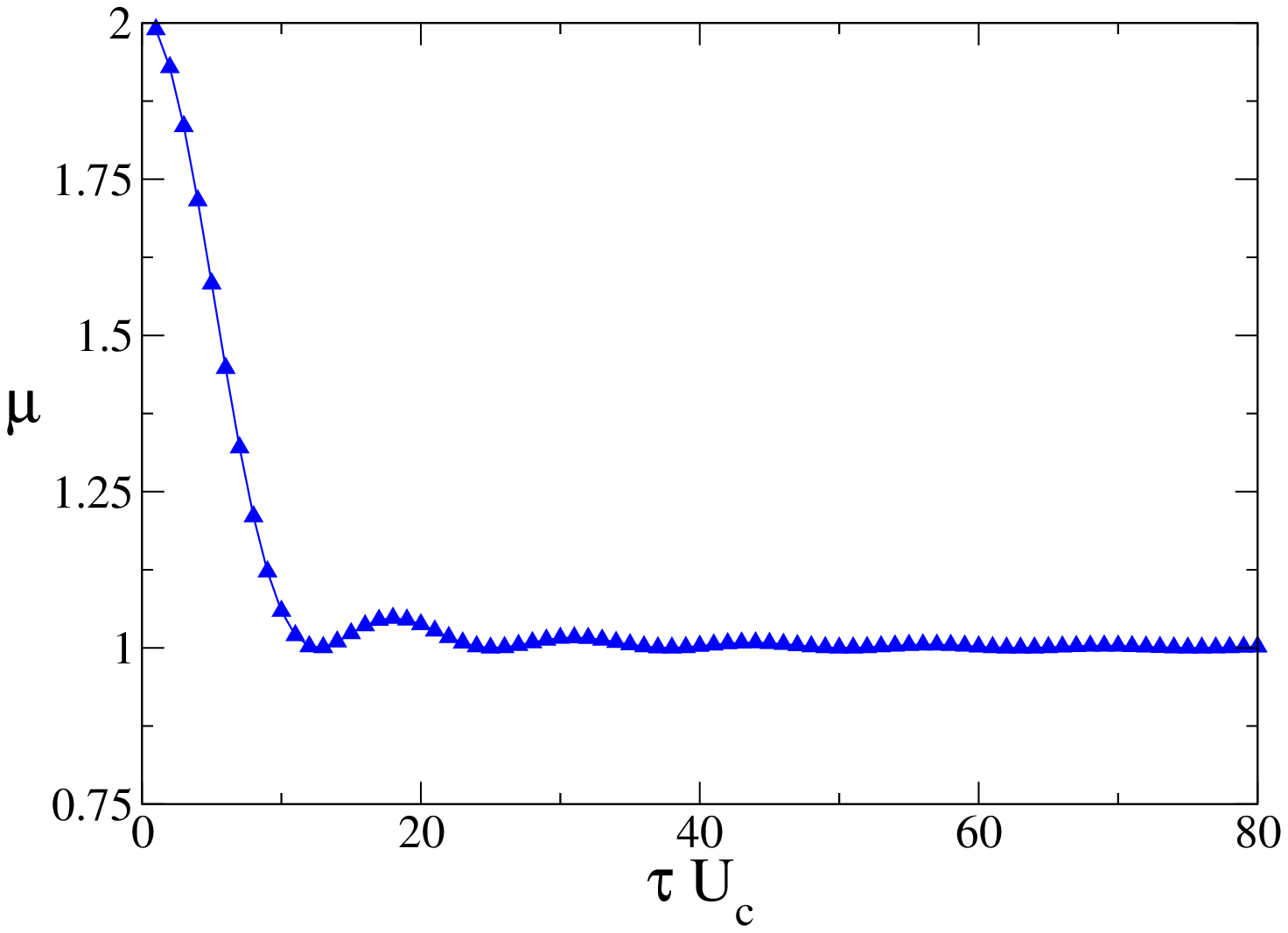,scale=0.31}
\caption{Mismatch $\mu(\tau)$, as defined in the main text, as a function of the ramp duration $\tau$. We see the crossover from the sudden quench limit $\mu=2$ to the adiabatic regime $\mu=1$. }
\label{fig:fig7}
\end{center}
\end{figure}
In light of this results we want now to understand how these metastable 
plateaux move as the ramp time is changed from the sudden to the adiabatic 
limit.

To this extent we compute the long time average of the quasiparticle weight, 
$\bar{Z}(u_f,\tau)$, and define, following Ref.~\onlinecite{Kehrein_ramps}, 
the mismatch function $\mu(\tau)$ as
\be
\mu(\tau)=\frac{1-\bar{Z}(u_f\rightarrow0,\tau)}{1-Z_{eq}(u_f\rightarrow0)}\,.
\ee
In the sudden quench case this quantity approaches $\mu(0^+)=2$, consistently 
with the results obtained with the flow equation method. In the opposite limit 
of a very long ramp we expect the mismatch to approach $\mu(\infty)=1$, namely 
the dynamics to be adiabatic. We plot in figure (\ref{fig:fig7}) the behaviour 
of the function $\mu(\tau)$, which shows a smooth crossover from the sudden to 
the adiabatic regime.

\section{Quantum Fluctuations Beyond Mean Field}\label{sect:sect4}

In this Section we discuss the role of quantum fluctuations 
on the mean field ramp dynamics we have previously presented. 
To this extent, we reformulate the Hubbard model in the framework 
of the $Z_2$ slave spin theory. We give here only the main results of 
this mapping and refer the reader to Ref.~\onlinecite{Z2-1,Z2-2,SchiroFabrizio_PRB11}.
As in other slave spin approaches~\cite{DeMedici-1,DeMedici-2}, we
map the local physical Hilbert space of the Hubbard Model onto 
the Hilbert space of an 
auxiliary spin model coupled to fermions and subject to a local constraint. 
In the $Z_2$ case the formulation is somehow minimal in the sense that 
auxiliary degrees of freedom are a single Ising spin and a spinful fermion.  
The Hamiltonian of the original Hubbard model, Eq.~(\ref{eqn:hubb}),
when written in terms of the auxiliary degrees of freedom reads
\bea
H_{Ising} &=& -t\sum_{<\bR,\bRp>\, \sigma}\, \sigma^x_\bR\,\sigma^x_\bRp\,
f^\dagger_{\bR\sigma}f^\dagga_{\bRp\sigma}\nonumber\\
&& + \frac{U(t)}{4}\sum_\bR\,\Big(1-\sigma^z_\bR\Big),\label{H-Ising-Hubbard}
\eea
where $f^\dagga_{\bR\sigma}, f^{\dagger}_{\bR\sigma}$ are  auxiliary fermionic fields 
while $ \sigma^x_\bR$ is an Ising spin variable. The  spin-fermion hamiltonian~(\ref{H-Ising-Hubbard}) lives in an enlarged Hilbert space containing on each site a spin-full fermion and an Ising variable.
In order to project onto the physical Hilbert space of the original Hubbard model one can introduce the following operator in any quantum average,
\be
\mathcal{Q} = \prod_\bR\left(\fract{1-\sigma^z_\bR\,\Omega_{\bR}}{2}\right),
\label{Z2-constraint-Hubbard}
\ee
where $\Omega_{\bR}=e^{i\pi\,n_{\bR}}$ and $n_{\bR}=\sum_{\sigma}\,f^{\dagger}_{\bR\sigma}\,f^{\dagga}_{\bR\sigma}$. The above operator
is actually a projector of the enlarged Hilbert space onto the subspace where 
if $n=1$ then $\sigma^z=+1$ while, if $n=0,2$, then $\sigma^z=-1$.
As a matter of fact, $\mathcal{Q}$ is just the constraint introduced in 
Ref.~\onlinecite{Z2-2} as a basis of the $Z2$ slave-spin 
representation of the Hubbard model. The constraint holds in general between Hilbert spaces, 
hence between evolution operators both in imaginary as well as in real time. 
This allows us to study the dynamics of the original Hubbard model using the 
Ising spin-fermion Hamiltonian ~(\ref{H-Ising-Hubbard}).

In Ref.~\onlinecite{SchiroFabrizio_PRB11} we have shown that (i) in 
infinite dimensions and at particle hole symmetry the constraint is 
ineffective and (ii)  that when gauge fluctuations are neglected, 
namely a product state between spins and fermions is assumed during the 
evolution, and when the resulting transverse field Ising model is treated in 
mean field, the time dependent Gutzwiller results follow.  
The advantage of this approach is that, once we have formulated the problem in the Ising 
language, we can attempt to include quantum fluctuations 
beyond mean field, even though this amounts to move away from infinite 
coordination lattices where the neglect of the constraint is not anymore 
justified.

This strategy was pursued in Ref.~\onlinecite{Z2-2} 
to study the zero temperature equilibrium Mott transition and then 
in Ref.~\onlinecite{SchiroFabrizio_PRB11} to access the dynamics 
after a sudden quench. Interestingly enough, this latter investigation 
revealed that quantum fluctuations become dynamically unstable in a region 
of quenches around the mean field critical line. Such a behaviour 
may be suggestive of an instability toward an inhomogeneous state where 
translational symmetry (which was implicitly assumed in the mean field dynamics)
is broken and may also suggest that the dynamical critical behaviour 
found at the mean field level gets strongly modified in finite dimensions. 

Here we would like to apply this mean field plus fluctuation 
approach to the problem of a finite ramp
and revisit in particular the analysis we have done in 
Section ~\ref{sect:sect3} on the scaling of excitation energy 
and the degree of adiabaticity of the process.  We expect that for 
sufficiently slow ramps, when the system stays close to the instantaneous 
ground state, no instability in the fluctuation spectrum should arise. This 
will allow to include gaussian fluctuations in a controlled way and to 
address questions concerning the Mott insulating state dynamics that 
otherwise are out of reach within the Gutzwiller mean field theory. 
Conversely, upon increasing the speed of the ramp, the simple treatment of 
fluctuations without feedback would again recover the unstable behaviour 
found in the sudden quench case. In order to go beyond this simple treatment, 
we develop here a self consistent treatment of quantum fluctuations and 
discuss the results of the coupled quantum-classical dynamics in the sudden 
quench limit.

\subsection{Fluctuations above mean field for slow ramps}

We start our discussion of fluctuations from the limit of very slow ramps. 
In this regime  when the dynamics is almost adiabatic, 
the fluctuations are expected to be well behaved, since in the limit of 
$\tau\rightarrow\infty$ we should recover the fluctuation spectrum in the 
instantaneous ground state of the Ising model which is known to be well 
behaved.~\cite{Z2-2} To this extent, we start from the 
Hamiltonian~(\ref{H-Ising-Hubbard}) and decouple the slave spins from the 
fermionic degrees of freedom, namely we assume a time dependent factorized wave function
$$
\vert\Phi(t)\rangle=|\Phi_s(t)\rangle\,|\Phi_f(t)\rangle
$$ 
each component $|\Phi_s(t)\rangle$ and $|\Phi_f(t)\rangle$ being translationally 
invariant. The electron wavefunction will evolve under the action of a time-dependent hopping, which 
is however still translationally invariant. Hence, if $|\Phi_f(t=0)\rangle$ is eigenstate of the 
hopping at $t<0$, in particular its ground state state, it will stay unchanged under the time evolution. 
Therefore we shall only focus on the evolution of the Ising component. Its effective Hamiltonian
$H_{s}=\langle\,H_{Ising}\rangle_{f}$ at positive times 
and in units of $U_c$ is 
\be
H_{s} = -\frac{u_f}{4}\sum_\bR\,\Big(1-\sigma^z_\bR\Big) - 
\frac{1}{8}\,\frac{2}{z}\sum_{<\bR\,\bRp>}\,\sigma^x_\bR\,\sigma^x_\bRp,\label{H-Ising-out}
\ee
We now follow the steps described in Ref.~\onlinecite{SchiroFabrizio_PRB11} and derive a time dependent
spin wave theory for the dynamics of this Ising model. We parametrize the dynamics generated by $H_{s}$ as a rotation of the spins, namely we choose a trial state in the spin sector of the form
$$
\vert\Phi_s(t)\rangle=\m{U}(t)\vert\Phi_{0}(t)\rangle
$$
where the unitary operator $\m{U}(t)$ 
$$
\m{U}(t)=e^{i\frac{\alpha}{2}\,\sum_\bR\,\sigma^x_\bR}\,
e^{i\frac{\beta}{2}\,\sum_\bR\,\sigma^y_\bR}\,
$$
defines a rotation of angles $\alpha,\beta$ which in general depend on time. By imposing the Schroedinger equation we conclude the state $\vert\Phi_{s0}(t)\rangle$ evolves with a transformed time dependent effective hamiltonian $H_{s\star}$ given by
$$
H_{s\star}(t)=-i\,\mathcal{U}(t)^\dagger\,\dot{\mathcal{U}}(t) + \mathcal{U}(t)^\dagger\,H_s\,\mathcal{U}(t)
$$
This effective hamiltonian can be treated within a spin-wave approximation in which the spin operators are expressed in terms of bosonic modes. We refer the reader to  Ref.~\onlinecite{SchiroFabrizio_PRB11} for further details. The dynamics for the angles $\alpha,\beta$ is obtained by requiring that the effective hamiltonian is quadratic in the bosonic operators. It is convenient to express the dynamics in terms of a different set of classical degrees of freedom, $\theta,\phi$ which are related to the angles $\alpha,\beta$ by 
\bea\label{eqn:alphabeta_thetaphi}
\cos\theta= \sin\beta\,\cos\alpha\\
\sin\theta\,\cos\phi= \cos\beta \\
\sin\theta\,\sin\phi = \sin\beta\,sin\alpha
\eea
The condition of vanishing linear terms gives~\cite{SchiroFabrizio_PRB11}
\bea\label{eqn:meanfield_thetaphi}
\dot{\theta} = \frac{1}{2}\,\sin\theta\,\cos\phi\,\sin\phi,\\
\dot{\phi} = -\frac{u(t)}{2}+\frac{1}{2}\,\cos\theta\,\cos^2\phi,
\eea
It is worth stressing that the above dynamics directly translates onto the Gutzwiller mean field dynamics (\ref{eqn:1D_dyn}) for the double occupancy $D(t)$ and its conjugate phase $\phi(t)$ upon posing
$D(t)=(1-\cos\theta)/4$ and $Z=\sin^2\theta\,\cos^2\phi$. While at the mean field level this is just an equivalent formulation, the slave spin framework allow to include quantum fluctuations which are lost in the Gutzwiller approximation. Indeed from the effective hamiltonian $H_{s\star}$ we have also access to the dynamics of fluctuations around the mean field trajectory described in terms of a quadratic time dependent bosonic Hamiltonian (see Ref.~\onlinecite{SchiroFabrizio_PRB11}). This reads \begin{figure}[t]
\begin{center}
\epsfig{figure=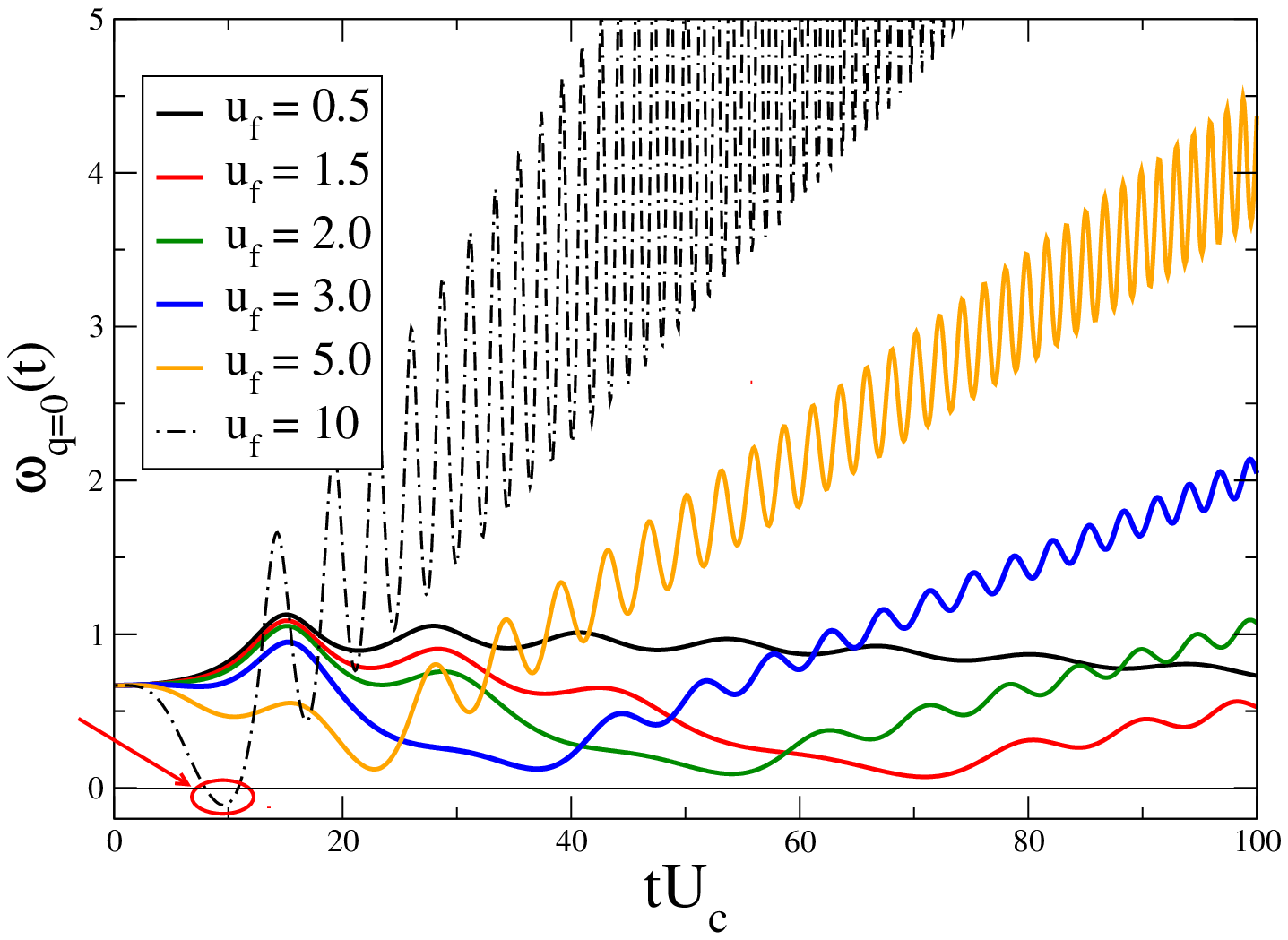,scale=0.31}
\caption{Evolution of fluctuation spectrum during a slow ramp ($\tau=100$). Conversely to the sudden quench case which showed an instability $(\omega^2_{\mathbf{q}=0}<0)$  we find that, 
beside a small window at short times and for large quenches, fluctuations are generally
well behaved.}
\label{fig:fig8}
\end{center}
\end{figure} 
%
\be
H_{qf}(t) = 
\sum_{\mathbf{q}}\,
\frac{1}{2m(t)}\,p_{\mathbf{q}}\,p_{-\mathbf{q}}+\frac{1}{2}\,m(t)\,
\omega^2_{\bq}(t)\,x_{\mathbf{q}}\,x_{-\mathbf{q}}
\ee
where the mass and the frequency read, respectively, as
\be\label{eqn:mass}
m(t)= \frac{2\,\left(1-\sin^2\theta\,\cos^2\phi\right)}{u(t)\,\cos\theta}
\ee
and
\be\label{eqn:spectrum}
\omega^2_{\bq}(t) = \left(\frac{u(t)\,\cos\theta}
{2\left(1-\sin^2\theta\,\cos^2\phi\right)}\right)^2
-\frac{u(t)}{4}\,\cos\theta\,\gamma_{\bq}
\ee
where, in a hypercubic lattice in $d$-dimensions,  
\[
\gamma_{\bq} = \frac{1}{d}\sum_{i=1}^d\,\cos q_i.
\]
We start discussing the behaviour of the excitation spectrum 
$\omega_{\mathbf{q}}(t)$ as a function of time for different values of the final 
interaction $u_f$ and for a slow ramp $\tau=100$. In 
Figure~\ref{fig:fig8} we plot in particular the value at $\bq=0$, 
which was found to be the most unstable mode in the sudden quench case.  
As we can see, except for very large quenches $u_f\gg1$ and short times, 
the spectrum is well behaved. In addition, from the structure of equations~(\ref{eqn:spectrum}
and the result obtained for the mean field dynamics, we conclude that for an infinitely slow ramp toward a final value of the interaction $u_f$ the out of equilibrium dynamics will be close to the instantaneous ground-state manifold, including the fluctuation contribution. 

Obviously a finite ramp time induces an excitation in the system and it is particularly interesting to see how the excitation 
energy $\Delta E_{exc}(\tau)$ scales to zero for very large $\tau$ and how the  spin wave spectrum affects this decay. To this extent we compute the total energy during the ramp 
$E(t)=\langle\Psi(t)\,\vert\,H(t)\,\vert\Psi(t)\rangle$ and get 
$\Delta E_{exc}(\tau)$ through the definition
\be
\Delta E_{exc}(\tau) = E(t=\tau) - E_{gs}(u_f)\,,
\ee
where the groundstate energy at the final value of the interaction can be computed within an equilibrium spinwave calculation and reads, for $u_f<1$
\be 
E_{gs}(u_f)=-\frac{1}{8}(1-u_f)^2-\frac{1}{4V}\,\sum_{\bq}\,\left(1-\sqrt{1-u_f^2\,\gamma_{\bq}}\right)
\ee
while in the Mott Insulating phase $u_f>1$
\be 
E_{gs}(u_f)=-\frac{u_f}{4V}\,\sum_{\bq}\,\left(1-\sqrt{1-\gamma_{\bq}/u_f}\right)
\ee
The total energy during the ramp is given by the kinetic and potential energy 
contributions
\be
E(t) = K(t)+u(t)\,D(t), 
\ee
which can be easily expressed as a mean field term plus a 
correction due to quantum fluctuation. In particular, 
we get for the double occupancy 
\bea
D(t) = \frac{1}{4}\left[1-\cos\theta
\left(\frac{1}{V}\,\sum_{\mathbf{q}}\,\left(1-\,\langle\,\Pi_{\mathbf{q}}\rangle_t\right)
\right)\right],
\eea
\begin{figure}[t]
\begin{center}
\epsfig{figure=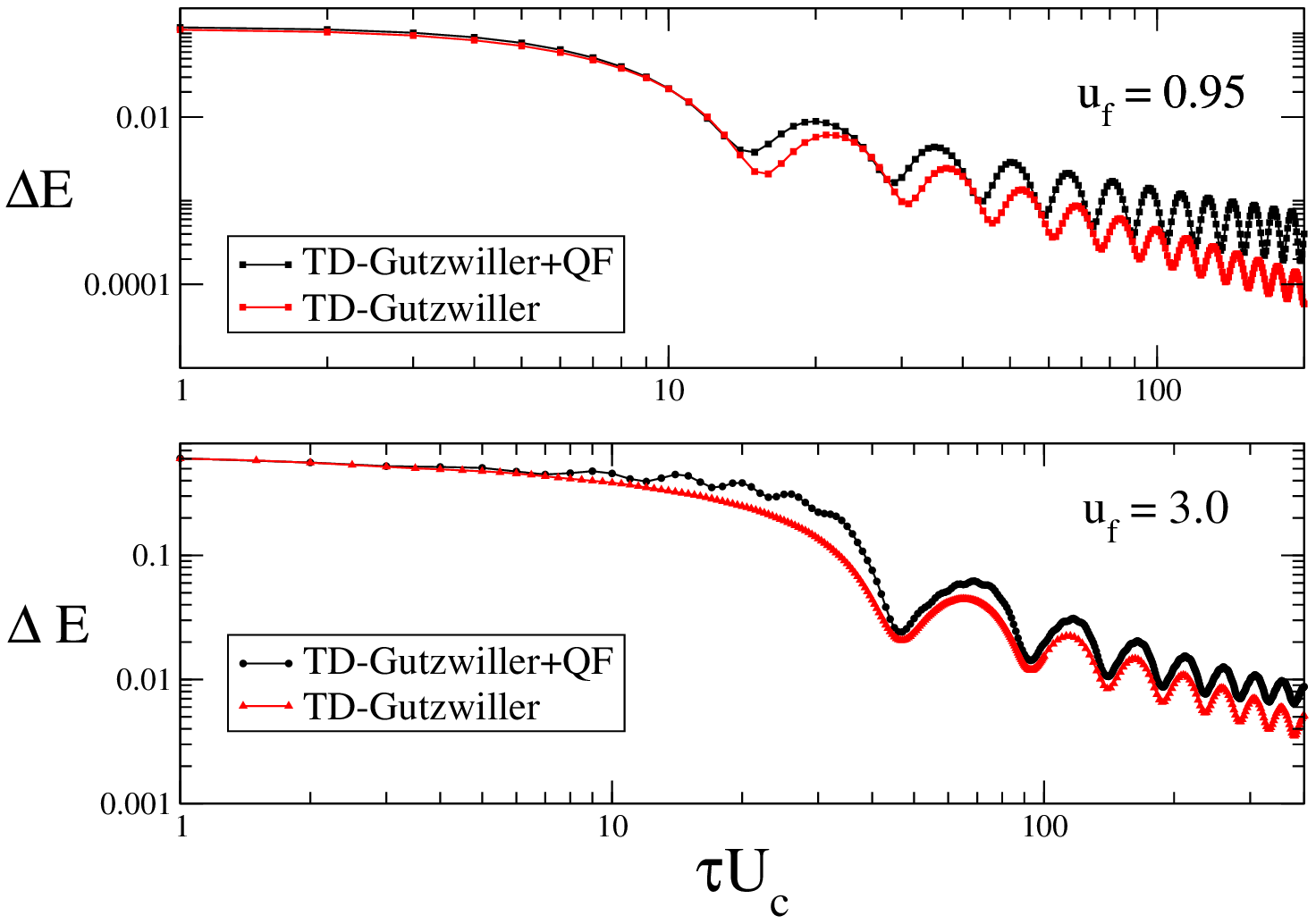,scale=0.31}
\caption{Excitation Energy $\Delta E(\tau)$ for ramps ending in the metallic  phase (top panel) or in the insulating  phase (bottom panel). We compare the results of time dependent Gutzwiller with those obtained by including quantum fluctuations at the gaussian level. We notice a sizeable effect of these  in the metallic case, which turns to be less pronounced for ramps ending in the insulating phase.}
\label{fig:exc_energyqf}
\end{center}
\end{figure}
while the kinetic energy (including both the coherent and the incoherent 
contribution) reads
\bea
K(t) &=& -\frac{1}{8V}\,\sin^2\theta\,\cos^2\phi\,\sum_{\mathbf{q}}\,
\left(1-2\,\langle\,\Pi_{\mathbf{q}}\rangle_t\right)+\\
&& -\frac{1}{4V}\,\left(1-\sin^2\theta\,\cos^2\phi\,\right)\,\sum_{\mathbf{q}}\,
\,\gamma_{\mathbf{q}}\,\langle\,x_{\mathbf{q}}\,x_{-\mathbf{q}}\rangle_t \nonumber 
\eea
where $\langle\,\Pi_{\mathbf{q}}\rangle_t$ measures 
the strength of quantum fluctuations and is defined by 
$$
\langle\,\Pi_{\mathbf{q}}\rangle = 
\langle\,x_{\mathbf{q}}\,x_{-\mathbf{q}}+p_{\mathbf{q}}\,p_{-\mathbf{q}}\rangle_t -1\,.
$$
It is useful at this point to write both the average energy $E(t)$ and its ground state value $E_{gs}(u_f)$ explicitly as a mean field part plus a correction due to quantum fluctuations. This allows us to disentangle the two contributions to the scaling of the excitation energy
\be
\Delta E_{exc}(\tau)=\Delta E_{exc}^{mf}(\tau)+\Delta\,E_{exc}^{qf}(\tau) 
\ee
where the mean field term has been discussed in previous sections while the quantum-fluctuations correction reads
\bea
\Delta E_{exc}^{qf}(\tau)=\frac{1}{4V}\,\sum_{\bq}\,\left(u_f\,\cos\theta+\sin^2\theta\,\cos^2\phi\right)\langle\Pi_{\bq}\rangle_{\tau}+\nonumber\\
-\frac{1}{4V}\,\sum_{\bq}\,\left(1-\sin^2\theta\,\cos^2\phi\right)\,\gamma_{\bq}\,\langle\,x_{\bq}\,x_{-\bq}\rangle_{\tau}\nonumber
\eea
We stress that the above quantum averages are taken over the dynamics generated by the time dependent hamiltonian $H_{qf}(t)$, which is solved numerically step by step together with the mean field dynamics~(\ref{eqn:meanfield_thetaphi}). To this extent we use a finite grid in 
momentum space (with typical size $N_{mesh}=100$) corresponding to a semielliptic density of states
\be\label{eqn:dos}
\rho(\eps) = \frac{2\sqrt{1-\eps^2}}{\pi}.
\ee
This makes the evaluation of $\Delta E^{qf}_{exc}$ a rather challenging numerical task, in particular for long ramp times where finite size effects become relevant and larger sizes are required to obtain converged results.  In figure~(\ref{fig:exc_energyqf}) we plot the behaviour of the excitation energy $\Delta E_{exc}(\tau)$ as a function of the ramp time $\tau$ starting from a non-interacting system and for final values of the interaction corresponding respectively to a metallic (top panel) and insulating (bottom panel) final state. The longest ramp time we were able to achieve, $\tau\sim 200$, although still not enough to obtain a robust scaling, allows us to attempt a discussion of the long time behaviour of $\Delta E_{exc}$ in presence of quantum fluctuations. As we see from figure~(\ref{fig:exc_energyqf}) the numerics suggest that quantum fluctuations do in fact affect the long-time behaviour of the excitation energy, particularly in the metallic phase and less strongly in the insulating phase. 

In order to rationalize this behaviour it is useful to resort to a more analitical approach. Indeed the hamiltonian of quantum fluctuations $H_{qf}(t)$ describe coupled harmonic oscillators with time dependent parameters (mass $m(t)$ and frequency $\omega_{q}(t)$). The characteristic time scale for their variation is given by the mean field dynamics of the variational parameters and from section~\ref{sect:sect3} we know we can describe this for long ramps as an adiabatic evolution plus small oscillations. Hence, using Eqs~(\ref{eqn:mass}-\ref{eqn:spectrum}) we can write $m(t)$ and $\omega_{\bq}(t)$ for large $\tau$ as
\bea
m(t)=m^{gs}(u(t/\tau)) +\frac{\delta m_{\tau}(t)}{\tau^{\delta}}\,\\
\omega_{\bq}(t)=\omega^{gs}_{\bq}(u(t/\tau))+\frac{\delta\omega_{\bq\tau}(t)}{\tau^{\delta}}\,,
\eea
where $\delta$ is a mean-field exponent that in general depends on the whether the ramp ends in the metallic or insulating phase,  while $\delta m_{\tau}(t),\delta\omega_{\bq\tau}(t)$ are pre-factors that can be computed from the mean field dynamics in the adiabatic limit (see appendix~\ref{sect:appA} for further details). This argument suggests that we can obtain the dynamics of quantum fluctuations as an expansion around the adiabatic limit~\cite{SotiriadisCardy_PRB}. In particular if we define
\be
\eta_{\bq}(t)=\frac{\dot{m}}{m}+\frac{\dot{\omega_{\bq}}}{\omega_{\bq}} 
\ee
we can obtain to leading order in $\eta_{\bq}$
\bea
\langle\,x_{\bq}\,x_{-\bq}\rangle_t = \frac{1}{2\,m(t)\omega_{\bq}(t)}\,
\left(1+\int_0^t\,dt'\,\cos\,2\theta_{\bq}(t,t')\,\eta_{\bq}(t')\right)\nonumber\\
\langle\,p_{\bq}\,p_{-\bq}\rangle_t = \frac{m(t)\omega_{\bq}(t)}{2}\,
\left(1-\int_0^t\,dt'\,\cos\,2\theta_{\bq}(t,t')\,\eta_{\bq}(t')\right)\nonumber
\eea
where 
$$
\theta_{\bq}(t,t')=\int_{t'}^t\,dt''\,\omega_{\bq}(t'')
$$
Using this result we can write the excitation energy due to quantum fluctuations $\Delta E^{qf}_{exc}$ as the sum of two contributions
\be\label{eqn:DeltaE_qf}
 \Delta E_{exc}^{qf}(\tau)=\Delta E_{exc}^{(1)}(\tau)+\Delta E_{exc}^{(2)}(\tau)
\ee
that read
\begin{widetext}
\bea
\Delta E_{exc}^{(1)}(\tau)=\frac{1}{4V}\sum_{\bq}\,\left[A(\tau)\left(\frac{m(\tau)\omega_{\bq}(\tau)}{2}+\frac{1}{2m(\tau)\omega_{\bq}(\tau)}-1\right)-\frac{\,B(\tau)\gamma_{\bq}\,}{2m(\tau)\omega_{\bq}(\tau)}\right]-E_{gs}^{qf}(u_f)\\
\Delta E_{exc}^{(2)}(\tau)=\frac{1}{4V}\sum_{\bq}\,\left[A(\tau)\left(\frac{1}{2m(\tau)\omega_{\bq}(\tau)}-\frac{m(\tau)\omega_{\bq}(\tau)}{2}\right)-\frac{\,B(\tau)\gamma_{\bq}}{2\,m(\tau)\omega_{\bq}(\tau)}\right]\int_0^{\tau}\,dt'\,\cos2\theta_{\bq}(\tau,t')\eta_{\bq}(t')
\eea
\end{widetext}
where the coefficients $A(\tau),B(\tau)$ read respectively as $A(\tau)=u_f\,\cos\theta(\tau)+\sin^2\theta(\tau)\cos^2\phi(\tau)$ and $B(\tau)=1-\sin^2\theta(\tau)\cos^2\phi(\tau)$. The two terms in Eq.~(\ref{eqn:DeltaE_qf}) have a clear interpretation as the first accounts for excitations produced by a non adiabatic mean field dynamics while assuming quantum fluctuations to follow adiabatically, while the second accounts for deviations from adiabaticity due to quantum fluctuations, with a mean field dynamics following its instantaneous ground state.
 Interestingly enough one can easily check that this latter contribution vanishes to leading order, (i.e. when $m(\tau)=m^{gs}(u_f)$ and $\omega_{\bq}(\tau)=\omega^{gs}_{\bq}(u_f)$) namely it only contributes to sub-leading order. The dominant contribution comes therefore from $\Delta E_{exc}^{(1)}(\tau)$ and quite generically would give rise to corrections of order $1/\tau^{\delta}$. While $\delta=1$ for ramps in the metallic phase and it is therefore a rather big correction to the mean field power law $\delta=2$, the situation is milder for ramps in the insulating side and this may explain the behaviour in figure~(\ref{fig:exc_energyqf}).


\subsection{Sudden Quench Limit: a self consistent theory of fluctuations}

In this Section we address the opposite limit of a sudden quench 
and formulate a self consistent theory of quantum fluctuations which goes 
beyond the previous treatment and that of Ref~\onlinecite{SchiroFabrizio_PRB11}. 
The crucial ingredient that we include here is the feedback of quantum 
fluctuations on the mean field dynamics which is expected to be relevant 
especially close to the mean field dynamical critical line 
where fluctuations would otherwise start to become unstable.
We give a detailed derivation of this new treatment of fluctuations in 
the Appendix~\ref{app:self_qf}. Here we briefly discuss the key features of 
this approach and the results of the quench dynamics. 
In order to couple the mean field dynamics and the fluctuations we took 
inspiration from the Bogoliubov theory of weakly interacting superfluids. 
There, a condensate classical order parameter is identified with the quantum 
degrees of freedom of modes at $\bq=0$ while those modes with $\bq\neq0$ 
represent the fluctuations out of the condensate. Assuming the 
classical order parameter to be a macroscopic one can simplify 
the commutation relations and get a closed set of equations of 
motion for the classical as well as the quantum components. In the case 
of present interest there is of course no real condensate as a discrete 
rather than continuous symmetry is broken in the quantum Ising model. 
However, we can still consider the modes at $\bq=0$, corresponding to the 
global magnetization, to be classical and macroscopic with the 
consequent simplification of the Heisenberg equations of motion for the 
modes at $\bq=0$ and  $\bq\neq0$. The resulting dynamics for the mean field 
part $\theta,\phi$ will read (see Appendix)
\bea\label{eqn:con_dyn}
\dot{\theta} &=& \frac{N}{2}\sin\theta\cos\phi\sin\phi \\
 & & +\frac{1}{2NV^2}\sum_{\bq \neq 0} \gamma_\bq \left( \sin\theta \De{xy} 
 + \cos\theta\sin\phi\De{xz} \right) \nonumber\\
\sin\theta \dot{\phi} &=& -\frac{u}{2}\sin\theta + \frac{N}{2}\sin\theta\cos\theta\cos^2\phi \nonumber \\
 & & + \frac{1}{2NV^2}\cos\phi\sum_{\bq \ne 0}\gamma_\bq\De{xz} \nonumber\\
\dot{N} &=& \frac{1}{2V^2}\sum_{\bq \neq 0}\gamma_\bq \left( -\cos\theta\De{xy} + \sin\theta\sin\phi\De{xz} \right) \nonumber
\eea
where $N(t)$ is the magnitude of the classical order parameter while 
$\Delta_{ab}(\bq,t)$ is a (time dependent) average for the modes with 
$\bq\neq0$ and it is defined as
\be
 \Delta_{ab}(\bq,t) \equiv \frac{1}{2}\aver{\s{a}{\bq}\s{b}{-\bq} 
+ \s{b}{\bq}\s{a}{-\bq}}_t\,\qquad
 a,b=x,y,z
\ee
The above dynamics differs from the conventional mean field Guztwiller dynamics introduced previously in two main respects. First, there is an explicit 
coupling of the modes at $\bq\neq0$ with the classical dynamics of 
$\theta,\phi$. Second, the amplitude $N$ of the order parameter is no more 
frozen but rather is allowed to change with time. The above dynamical 
system can be closed by writing the equation of motion for 
$\Delta_{ab}(\bq,t)$. The result takes the form (see Appendix~\ref{app:self_qf})
\be
\partial_t\, \Delta_{ab}(\bq,t) = \sum_{cd}\,M_{abcd}(\bq)\,\Delta_{cd}(\bq,t)
\ee
where the coefficients $M_{abcd}(\bq)$ depend in general from 
both $\theta,\phi$ and $N$. As we show in the Appendix, 
the above dynamics conserves the total energy of the system after the quench, 
a crucial feature that was missing in the spin-wave treatment of fluctuations. 
\begin{figure}[t]
\begin{center}
\epsfig{figure=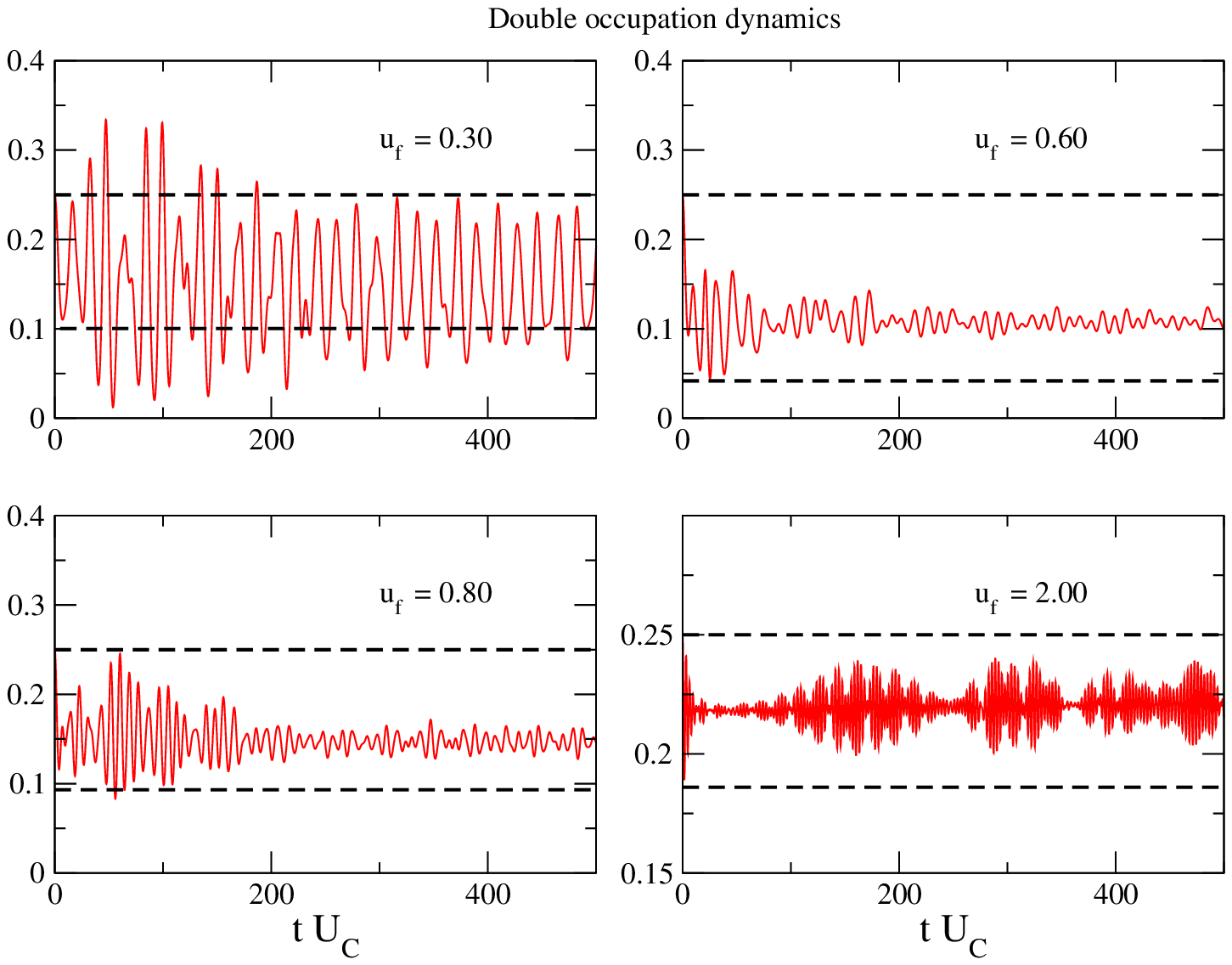,scale=0.35}
\vspace{10pt}
\epsfig{figure=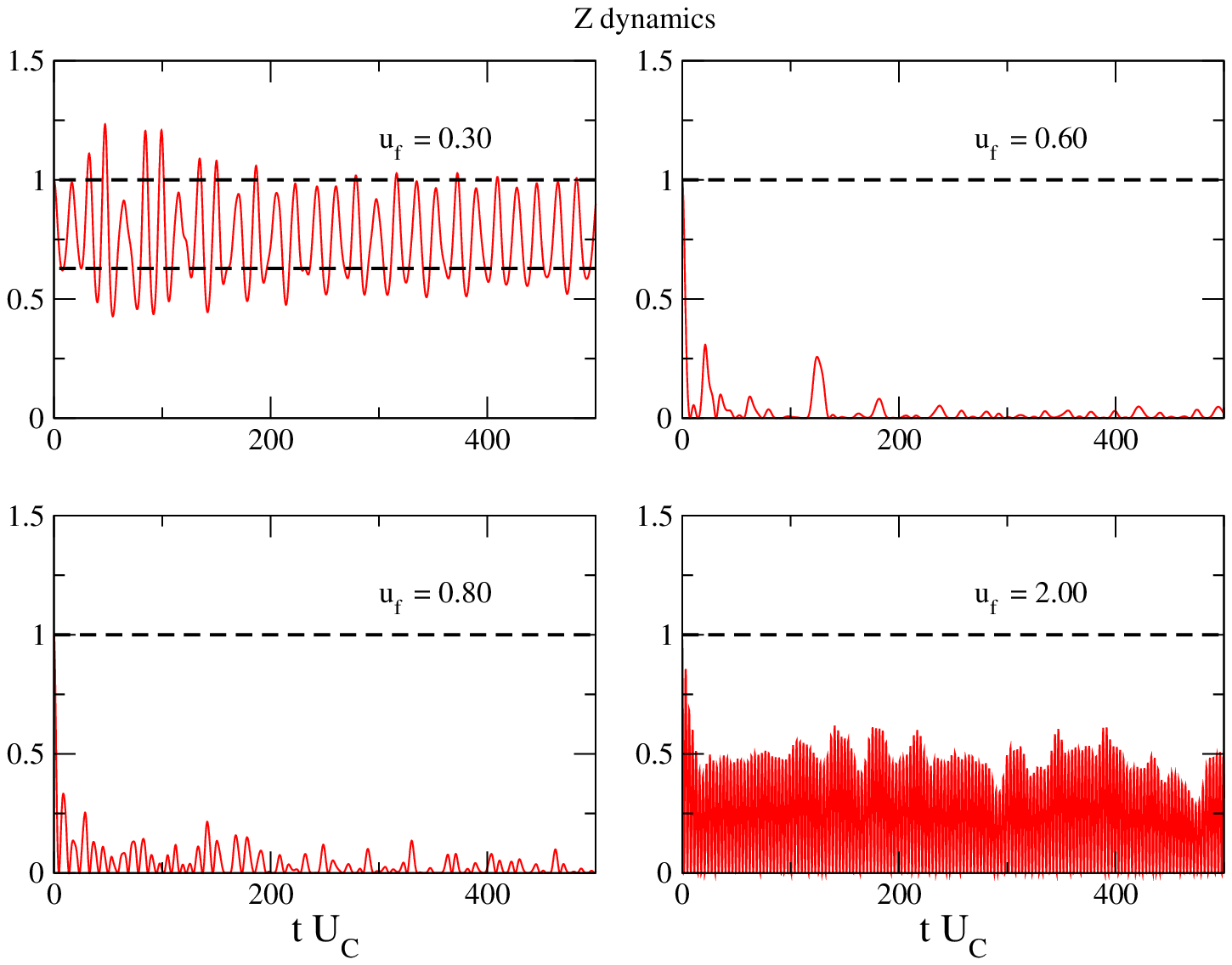,scale=0.35}
\caption{Time dynamics of the double occupation $D(t)$ (top panel) and of the quasiparticle weight $Z(t)$ (bottom panel). The dashed black lines are guide to eyes and bound the region where the mean field dynamics with no quantum fluctuations would display coherent oscillations. In the last three panels oscillations are between $0$ and $1$. The red curves are the dynamics obtained from the numerical solution of (\ref{eqn:con_dyn}-\ref{eqn:fluc_dyn}).}
\label{fig:D_Dyn1}
\end{center}
\end{figure}
We now discuss the numerical solution of the above coupled dynamics for a 
quantum quench from a non interacting initial state. As in the previous section 
in order to solve numerically the coupled dynamics we use a finite grid in 
momentum space corresponding to the semielliptic density of states~(\ref{eqn:dos}).
We expect that in the region where fluctuations are negligible, 
the time dependent spin wave approximation is recovered and the system 
will display an oscillatory dynamics with multiple frequencies but no 
real damping. As opposite, close to the critical region where 
fluctuations become important and spin wave approximation breaks down, 
we expect the feedback of the modes at $\bq\neq0$ on the classical dynamics 
to be extremely relevant in setting the steady state.
\begin{figure}[t]
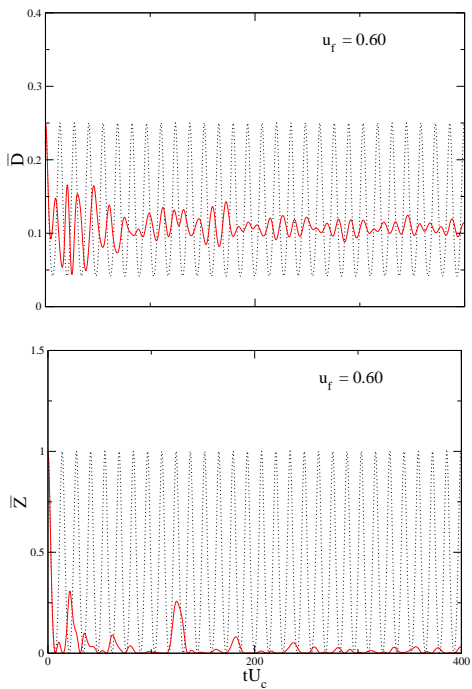

\begin{center}
\epsfig{figure=fig_newD.eps,scale=0.25}
\vspace{10pt}
\epsfig{figure=fig_newZ.eps,scale=0.25}
\caption{Short time dynamics with the feedback of quantum fluctuation for double occupation and quasiparticle weight and comparison with the mean field dynamics. We see that the coherent oscillations present at the mean field level are quickly damped out as the contribution of quantum fluctuations is properly taken into account.}
\label{fig:D_Dyn2}
\end{center}
\end{figure}
In Fig.~\ref{fig:D_Dyn1}-\ref{fig:D_Dyn2} we plot the time dynamics of the double occupation, 
$D(t)$, and that of the quasiparticle weight, $Z(t)$, 
for different values of $u_f$. To enlight the effect of quantum fluctuations, 
in the same figure we bound with two dashed lines 
the region where the mean field dynamics for $D(t)$ and $Z(t)$ would display simple oscillations.
As expected, we note that approaching the critical region 
the coupling of fluctuation with the classical sector tends to drive the 
dynamics of local observables towards stationarity. In particular, one can 
see that for quenches ranging in a window from $u_f\approx 0.35$ to 
$u_f\approx0.9$, the dynamics of $D(t)$ and $Z(t)$ is quickly damped, 
while for smaller and larger values of $u_f$ fluctuations are less effective 
in driving the dynamics toward a steady state and some undamped oscillations 
are still clearly visible.

Another interesting feature emerging from the solution of the coupled dynamics 
concerns the fate of the mean field dynamical critical point upon including 
the feedback of fluctuations. This issue was not fully addressed in 
Ref.~\onlinecite{SchiroFabrizio_PRB11}, since 
the spin wave approach breaks down before the critical point 
due to the instability of quantum fluctuations. The solution of 
the coupled classical-quantum dynamics reveals that a kind of 
dynamical transition is still present even with quantum fluctuations. 
This is evident if one looks at the dynamics of the phase $\phi$, 
conjugate to the double occupation $D(t)$. Indeed, such a quantity 
still features a sharp pendulum-like dynamical instability at a 
finite value of the interaction $u_f^c$ which now gets modified by 
fluctuations and renormalized toward a smaller value $u_{f, QF}^c\simeq0.35$ 
to be compared with the mean field estimate $u_{f, MF}^c\simeq0.5$. 

Finally it is interesting to discuss the behaviour of the long time averages 
$\bar{D},\bar{Z}$ as a function of $u_f$. At the mean field level, 
those averages contain a clear signature of the dynamical critical point as 
the special point at which both $D(t)$ and $Z(t)$ relaxes toward zero. 

\begin{figure}[!t]
\begin{center}
\epsfig{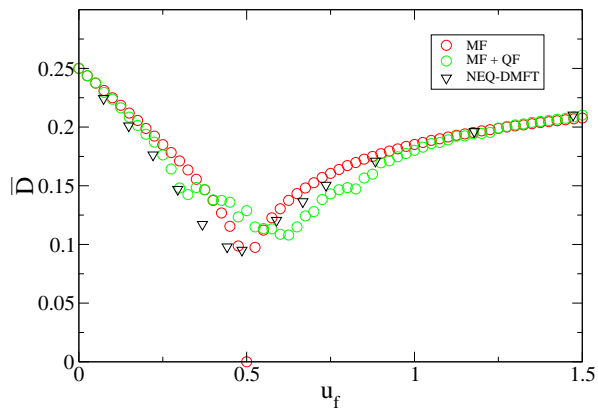}
\caption{Long-time average of $D(t)$; one can see that in vicinity of the critical region the inclusion of fluctuations corrects the mean field result. Instead, at weak and large values of $u_f$ the dynamics resembles the mean field one.}
\label{fig:confronto_D}
\end{center}
\end{figure}

\begin{figure}[!t]
\begin{center}
\epsfig{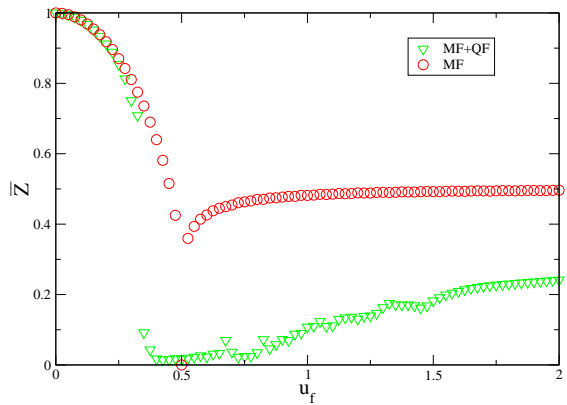}
\caption{Long-time average of $Z(t)$. With the inclusion of fluctuations the quasiparticle weight rapidly approaches zero in a region around the dynamical critical point. Such a behaviour is only partially catched by mean field dynamics.}
\label{fig:confronto_Z}
\end{center}
\end{figure}

In Fig.~\ref{fig:confronto_D} and \ref{fig:confronto_Z} we plot the 
behaviour of these long time averages with respect to $u_f$ and compare 
the respective time averages in the mean field dynamics and 
the results of out-of-equilibrium DMFT~\cite{Werner_prl09}. The result of this 
comparison seems to be consistent with the analysis of the transient dynamics 
and reveals the presence of three different regimes.
For weak quenches, quantum fluctuations do not play a major role and we 
recover almost exactly the mean field result. In this regime, time averages 
capture those predicted by perturbation theory\cite{Werner_prl09} for a 
pre-thermal state: $\bar{D}$ tends to the zero-temperature equilibrium value 
and 
$$
\bar{Z} \approx (2Z_{eq} -1 ).
$$
For quenches that approach the dynamical critical point, which in mean field 
dynamics corresponds to $u_f = 0.5$, we already saw that the dynamics of 
$D(t)$ and $Z(t)$ is rapidly driven towards a stationary state; $\bar{Z}$ 
maintains almost a constant zero value in this interaction window so that 
it shows a sharp variation with respect to the mean field value. 
Also $\bar{D}$ corrects the mean field result which was equal to zero at 
the dynamical critical point. Finally, for values of $u_f \gtrsim 0.9$, 
no fast relaxation occurs in the dynamics and time averages recover the mean 
field results, at least for the double occupation. The coherent part of the 
kinetic energy gets strongly suppressed with respect to the mean field 
average, a result that can be understood as due to a transfer of weight 
to the incoherent modes which are absence at the level of Gutzwiller. 
Overall, we could say that, upon including the feedback of quantum 
fluctuations on top of the mean field dynamics, we obtain a picture for the 
dynamics which is in substantial agreement with DMFT results.

\section{Conclusion}\label {sect:sect5}

In this paper we have discussed the non equilibrium dynamics of the fermionic 
Hubbard model after a linear ramp of the interaction $U$ across the Mott 
transition, starting from the metallic side. Our results are based on a time 
dependent Gutzwiller variational approach and on a theory of mean field plus 
fluctuations that we have developed in the framework of the $Z_2$ slave 
spin approach.
We have discussed the dynamics during the ramp and the issue of adiabaticity 
of the protocol by computing the excitation energy and studying its scaling 
for long ramp times. In addition we have discussed the dynamics after the 
ramp is completed, namely on time scales longer than $\tau$, and identified a 
dynamical transition at the mean field level which smoothly connects with 
the one already discussed for the sudden quench case. The properties of this 
transition as a function of the ramp time have been analyzed. Finally we have 
discussed the role of fluctuations on top of the mean field dynamics for both 
regimes of slow and very fast ramps. In the former case a gaussian theory of 
fluctuations is sufficient as the spectrum of the fluctuating modes is always 
well defined. Using this gaussian theory we have calculated the scaling of 
the excitation energy with $\tau$ and see how this is affected by the presence 
of a non trivial spectrum. An interesting extension of this kind of approach 
could be to look at the evolution of the spectral function in order to 
understand where the excitation energy due to the ramp is mostly  transferred. 
For what concerns short ramps we have developed a self consistent treatment 
of quantum fluctuations that goes beyond the simple gaussian theory. By means 
of this novel approach we have been able to obtain a finite and sizeable 
damping and the relaxation to a steady state, at least in the region close 
to the critical point.

\textit{Acknowledgements - } We would like to thank G. Biroli, C. Castellani and particularly G. Semerjian  for interesting discussions related to this work. This work has been supported by the European Union,
Seventh Framework Programme, under the project GO FAST, grant agreement No. 280555.
\appendix

\section{Classical Adiabatic Dynamics and Slow Quantum Fluctuations for long ramps}\label{sect:appA}

Here we discuss the classical dynamics of the Gutzwiller variational parameters in the limit of slow ramps and the small quantum fluctuations around it. We start from the equations of motions
\bea\label{eqn:1D_dyn}
\dot{D} = \frac{\bar{\eps}}{2}\,\fract{\partial Z}{\partial \phi}\\
\dot{\phi} = \frac{U(t)}{2} - \frac{\bar{\eps}}{2}\,\frac{\partial Z}{\partial D}
\eea
where $\bar{\eps}=\frac{U_c}{8}$ is the kinetic energy of the Fermi Sea in 
units of the critical repulsion $U_c$ for the zero temperature equilibrium Mott transition, 
while $Z[D,\phi] = 8D\,\left(1-2D\right)\,\cos^2\phi$ is the time dependent quasiparticle weight at half-filling. The above dynamics derives from a classical hamiltonian which reads
\be
E[D,\phi]=\frac{U(t)}{2}D-\frac{\bar{\eps}}{2}\,Z[D,\phi]
\ee
When $U(t)\equiv U_f/U_c=u_f\le 1$ the equilibrium solution 
\be
D_{gs}=\frac{1-u_f}{4}\,\qquad\,\phi_{gs}=0
\ee
is a stationary point of the hamiltonian. For slow variations of $U(t)$, that is for $\tau\rightarrow\infty$, we can assume to leading order the trajectory $D,\phi$ to follow the instantaneous minimum $D_{gs}(u(t/\tau)),0$ plus small oscillations that we want to compute. To this extent we expand $E$ around $D_{gs},\phi_{gs}$ up to the quadratic order (the first non vanishing). The result takes the form
\be
E=E_{gs}+ \frac{1}{2m(s)}\,\phi^2+\frac{1}{2}\,m(s)\,\omega^2(s)\,(D-D_{gs}(s))^2
\ee
where we have introduced $s=t/\tau$ as well as the slowly varying mass and frequency which read 
\be
m(s)=\frac{8}{1-(u_f\,s)^2}\,\qquad\,
\omega(s) =\frac{1}{2}\,\sqrt{1-(u_fs)^2}
\ee
We notice that for $u_f<1$ the frequency is always positive definite, while for ramps that cross the critical points it exists a time $t^{\star}=\tau/u_f<\tau$ at which the harmonic approximation breaks down. Let's first consider the case $u_f<1$. Then using results from classical adiabatic dynamics we can write to leading order
\be
D_{\tau}(s)=D_{gs}(s)-\frac{1}{\tau}D^{'}_{\star}(0)\sqrt{\frac{m(0)}{m(s)\omega(s)\omega(0)}}\,\sin\left(\tau\Omega(s)\right)
\ee
and 
\be
\phi_{\tau}(s)=\frac{m(s)\,D^{'}_{\star}(s)}{\tau}
-\frac{D^{'}_{\star}(0)}{\tau}
\sqrt{\frac{m(s)\,m(0)\omega(s)}{\omega(0)}}\,\cos\left(\tau\Omega(s)\right)\\
\ee
where we have defined $\Omega(s)=\int_0^s\,ds'\,\omega(s')$
\be
\Omega(s)=\frac{1}{4u_f}\left(u_fs\,\sqrt{1-(u_fs)^2}+\arcsin(u_fs)\right)
\ee
After simple algebra we get for 
\be 
D_{\tau}(s)=D_{gs}(s)+\frac{u_f}{2\tau}\left(1-(u_fs)^2\right)^{1/4}\,\sin\,\Omega(s)\tau
\ee
as well as
\be 
\phi_{\tau}(s)=\frac{1}{\tau}
\left(
-\frac{2u_f}{1-(u_fs)^2}+\frac{2u_f}{\left(1-(u_fs)^2\right)^{1/4}}\,\cos\Omega(s)\tau
\right)
\ee
The excitation energy $\Delta E(\tau)=E(t=\tau)-E_{gs}(u_f)$ can be easily evaluated in terms of $D_{\tau}(s=1)$ and $\phi_{\tau}(s=1)$ and the result gives the scaling $\Delta E_{exc}\sim 1/\tau^2$ quoted in the main text. As we mentioned, in the case of a ramp crossing the critical point, $u_f>1$, the situation is different as it exists a value $s_{\star}=1/u_f<1$ at which the frequency of small oscillations vanishes. The way to get around this and obtain the scaling of $D_{\tau}(s),\phi_{\tau}(s)$ for $s\in(s_{\star},1)$, discussed for example in Ref.~\onlinecite{Guilhem_ramps,Guilhem_privatecomm}, is to expand the hamiltonian around its bifurcation point, $(s_{\star},D_{\star}=0,\phi_{\star}=0)$The result reads
\be
E=E_{\star}+\frac{u_f}{2}(s-s_{\star})\,D+D^2+\frac{u_f}{8}(s_{\star}-s)\,\phi^2+\frac{1}{6}\phi^2\,D\nonumber 
\ee
and the classical dynamics for $\phi$ can be written as
\be
\ddot{\phi}=\frac{4}{6}u_f(s-s_{\star})\phi+\frac{1}{16}\phi^3+\frac{u_f}{2\tau}
\ee
which can be cast after a proper rescaling of variables into the form of a Painleve equation of second type. This treatment allows us to extract the scaling exponent of $\phi,D$ in the regime of large $\tau$
that gives respectively $D\sim \delta D/\tau^{2/3}$ and $\phi\sim\delta \phi/\tau^{1/3}$.

Let's now consider the effect of harmonic quantum fluctuations (QF) around the Gutzwiller dynamics in the regime of slow ramps~\cite{SotiriadisCardy_PRB}. The hamiltonian of QF describe a set of harmonic oscillators with time dependent mass $m(t)$ and frequency $\omega_{\bq}(t)$. In the limit of slow ramps, using the results just obtained for the mean field variational parameters and the definition of $m(t),\omega_{\bq}(t)$ in terms of $\theta,\phi$, we can write
\bea
m(t)=m^{gs}(u(t/\tau)) +\frac{\delta m_{\tau}(t)}{\tau^{\delta}}\,\\
\omega_{\bq}(t)=\omega^{gs}_{\bq}(u(t/\tau))+\frac{\delta\omega_{\bq\tau}(t)}{\tau^{\delta}}\,.
\eea
The exponent $\delta$ depends on whether the ramp ends below or above the critical point and from the previous discussion reads $\delta=1$ for $u_f<1$ and $\delta=2/3$ for $u_f>1$. The pre factors can in principle also computed, by a straightforward expansion in the case $u_f<1$ and by a slightly more complicated analysis for $u_f>1$ that requires a proper matching of scaling functions between the two regimes $s<s_{\star}$ and $s>s_{\star}$. Since we don't need these factors for our actual purpose here we will not discuss this point further.

In order to discuss the dynamics of quantum fluctuations in the limit of slow ramps,  we will for simplicity drop the index $\bq$ since, at the gaussian level we are considering here each mode evolves independently. Hence, considering just a single mode we have
\be
H(t) = 
\frac{p^2}{2m(t)}\,+\frac{1}{2}\,m(t)\,
\omega^2(t)\,x^2
\ee
Let's define the (explicitly time-dependent) annihilation/creation operators as
\bea\label{eqn:a_adagger}
a = \sqrt{\frac{m(t)\omega(t)}{2}}\,x-i\sqrt{\frac{1}{2m(t)\omega(t)}}\,p\\
a^{\dagger} = \sqrt{\frac{m(t)\omega(t)}{2}}\,x+i\sqrt{\frac{1}{2m(t)\omega(t)}}\,p
\eea
which satisfy the coupled equations
\bea
\dot{a}=-i\omega\,a+\frac{1}{2}\eta(t)\,a^{\dagger}\\
\dot{a}^{\dagger}=i\omega\,a^{\dagger}+\frac{1}{2}\eta(t)\,a^{\dagger}
\eea
with  $\eta=\partial_t\left(log\,m\,\omega\right)$. While a formal solution of this equations can be written in terms of time-ordered exponential we can perturbative expand the equation in power of $\eta$, the leading order term reading
\bea
a(t)=e^{-i\int_0^t\,\omega(t')dt'}\,a(0)+\nonumber\\
+e^{-i\int_0^t\,\omega(t')dt'}\,\int_0^t\,dt'\,e^{2i\int_0^{t'}\,dt''\omega(t'')}\,\eta(t')\,a^{\dagger}(0)
\eea
Using this result and the definitions~(\ref{eqn:a_adagger}) we can easily obtain the expression for coordinate and momentum operators, $x(t), p(t)$ in term of their initial values. Then, assuming the initial state to be in the ground state of $H(t=0)$ we can obtain the results quoted in the main text for $\langle\,x^2\rangle_t,\langle\,p^2\rangle_t$.

\section{Quantum Fluctuations plus feedback}\label{app:self_qf}

In this Appendix we present a treatment of quantum fluctuations above the 
mean field dynamics that goes beyond the spin wave (gaussian) approximation 
of Ref.~\onlinecite{SchiroFabrizio_PRB11} and that leads to the 
the dynamical equations (\ref{eqn:con_dyn}-\ref{eqn:fluc_dyn})  
we used in the main text.\\

As we showed in \onlinecite{SchiroFabrizio_PRB11}, in the approximation that 
the evolving state is a product of fermions and spins wavefunctions, the 
Hamiltonian (\ref{H-Ising-Hubbard}), upon introducing the Fourier transform of 
the Ising spins 
\ba
 \s{a}{\bq} = \sum_\bR e^{-i\bq\cdot\bR} \s{a}{\bR}
\ea
reads (in units of $U_c$)
\be
 \m{H}_I = \frac{u}{4}\s{z}{0} - \frac{1}{8V}\s{x}{0}\s{x}{0} - \frac{1}{8V}\sum_{\bq \ne 0}\gamma_\bq \s{x}{\bq}\s{x}{-\bq}.
\ee
$V$ is the number of sites and $\gamma_\bq= \sum_\mathbf{a} e^{i\bq\mathbf{a}}$, with $\mathbf{a}$ a vector which connects two nearest neighbor sites. The spin operators in momentum space satisfy the commutation relations
\be 
 [\s{a}{\bq},\s{b}{-\bq'}] = 2i \eps_{abc}\s{c}{\bq-\bqp}.
\ee
In the same spirit of the spin-wave approximation we assume that the evolved state has a \textit{condensate} component, which means that $\aver{\s{a}{0}} \sim V$ while, for any $\bq \neq 0$, $\aver{\s{a}{\bq}} =0$ because of translational symmetry and $\aver{\s{a}{\bq}\s{b}{-\bq}} \sim V$. If the dynamics is able to drive the system towards equilibrium, we expect a damping of the $\bq=0$ sector.\\
Within such an approach and at the leading order in $V$, the only non-vanishing commutation relations are
\ba
 \left[\s{a}{0},\s{b}{0} \right] &=& 2i\eps_{abc}\s{c}{0}\\
 \left[\s{a}{\bq},\s{b}{-\bq} \right] &=& 2i\eps_{abc}\s{c}{0}\\
 \left[\s{a}{\bq},\s{b}{0} \right] &=& 2i\eps_{abc}\s{c}{\bq} .
\ea
We evaluate then the equations of motions
\ba 
i\partial_t \s{a}{\bq} = \left[ \s{a}{\bq},\m{H}_I \right]
\ea
using the above approximate commutators. We find
\bea\label{eqn:ke0}
i\partial_t \s{x}{0} &=& i\frac{u}{2} \s{y}{0}\\
i\partial_t \s{y}{0} &=& -i\frac{u}{2} \s{x}{0} + \frac{i}{4V} \left(
			\s{z}{0}\s{x}{0} + h.c. \right) \nonumber \\
 		     & & +\frac{i}{4V}\sum_{\bq\ne0} \gamma_\bq \left(
			\s{z}{\bq}\s{x}{-\bq} + \s{x}{\bq}\s{z}{-\bq} \right)
			 \nonumber\\
i\partial_t \s{z}{0} &=& -\frac{i}{4V} \left(
			\s{x}{0}\s{y}{0} + h.c. \right) \nonumber \\
 		     & & -\frac{i}{4V}\sum_{\bq\ne0} \gamma_\bq \left(
			\s{x}{\bq}\s{y}{-\bq} + \s{y}{\bq}\s{x}{-\bq} \right)
			 \nonumber
\eea
for the $\bq=0$ components, while for the $\bq \neq 0$ ones
\bea\label{eqn:kd0}
i\partial_t \s{x}{\bq} &=& i\frac{u}{2} \s{y}{\bq}\\
i\partial_t \s{y}{\bq} &=& -i\frac{u}{2} \s{x}{\bq} + \frac{i}{2V} 
			\s{z}{\bq}\s{x}{0} +\frac{i}{2V}\gamma_\bq \s{x}{\bq}\s{z}{0} \nonumber\\
i\partial_t \s{z}{\bq} &=& -\frac{i}{2V} \s{y}{\bq}\s{x}{0} 
 		           -\frac{i}{2V} \gamma_\bq 
			   \s{x}{\bq}\s{y}{0} \nonumber.
\eea
We let then evolve the condensate component as a mean field, i.e. we assume for the $\bq=0$ spins the classical values
\bea
\s{x}{0} &=& VN\sin\theta\cos\phi\\
\s{y}{0} &=& VN\sin\theta\sin\phi \nonumber\\
\s{z}{0} &=& VN\cos\theta \nonumber
\eea
while for the $\bq \neq 0$ we introduce the following quantity
\be
 \Delta_{ab}(\bq,t) \equiv \frac{1}{2}\aver{\s{a}{\bq}\s{b}{-\bq} + \s{b}{\bq}\s{a}{-\bq}}.
\ee
From eq. (\ref{eqn:ke0}) and (\ref{eqn:kd0}) the dynamics of these quantities is easily derived and amounts to a set of non-linear coupled differential equations; the condensate dynamics satisfies
\bea\label{eqn:con_dyn}
\dot{\theta} &=& \frac{N}{2}\sin\theta\cos\phi\sin\phi \\
 & & +\frac{1}{2NV^2}\sum_{\bq \neq 0} \gamma_\bq \left( \sin\theta \De{xy} 
 + \cos\theta\sin\phi\De{xz} \right) \nonumber\\
\sin\theta \dot{\phi} &=& -\frac{u}{2}\sin\theta + \frac{N}{2}\sin\theta\cos\theta\cos^2\phi \nonumber \\
 & & + \frac{1}{2NV^2}\cos\phi\sum_{\bq \ne 0}\gamma_\bq\De{xz} \nonumber\\
\dot{N} &=& \frac{1}{2V^2}\sum_{\bq \neq 0}\gamma_\bq \left( -\cos\theta\De{xy} + \sin\theta\sin\phi\De{xz} \right) \nonumber
\eea
while the $\bq \neq 0$ terms
\bea\label{eqn:fluc_dyn}
\DDe{xx}&=& u\De{xy} \\
\DDe{xy}&=& \frac{1}{2}\left( -u + N\gamma_\bq\cos\theta \right)\De{xx} \nonumber\\
        & & +\frac{N}{2}\sin\theta\cos\phi\De{xz} + \frac{u}{2}\De{yy}
\nonumber\\
\DDe{xz}&=& -\frac{N}{2}\gamma_\bq\sin\theta\sin\phi\De{xx} 
\nonumber\\
        & & -\frac{N}{2}\sin\theta\cos\phi\De{xy} + \frac{u}{2}\De{yz}
\nonumber \\
\DDe{yy}&=& \left(-u + N\gamma_\bq\cos\theta \right)\De{xy} + N\sin\theta\cos\phi\De{yz} 
\nonumber \\
\DDe{yz}&=&-\frac{N}{2}\gamma_\bq\sin\theta\sin\phi\De{xy} 
\nonumber \\
 & & +\frac{1}{2}\left(-u + N\gamma_\bq\cos\theta\right)\De{xz} 
\nonumber \\
 & & -\frac{N}{2}\sin\theta\cos\phi\De{yy}+\frac{N}{2}\sin\theta\cos\phi\De{zz}
\nonumber \\
\DDe{zz}&=& -N\gamma_\bq\sin\theta\sin\phi\De{xz} - N\sin\theta\cos\phi\De{yz}
\nonumber
\eea
By inspection of (\ref{eqn:con_dyn}) one recognizes that if the feedback of the $\bq\neq0$ terms is neglected, the condensate dynamics is the same we obtained in the Gutzwiller approximation. In that approach indeed, $N$ remained fixed during the dynamics ($N(t)=1$), so that no damping was present for the condensate sector with a consequent impossibility of energy conservation. 
With respect to the results of Ref.~\onlinecite{SchiroFabrizio_PRB11}, this new approach has the main advantage to conserve the mean value of energy during the dynamics,
\ba 
\partial_t \aver{\m{H}} = 0
\ea
as one can easily verify from eq (\ref{eqn:con_dyn}-\ref{eqn:fluc_dyn}).\\
In this work we considered quenches from the non-interacting system ($u_i=0$); the initial conditions are then readily found from the solution of an Ising model in absence of transverse field and read:
\be
\left \{
\begin{array}{cc}
N(0)=&1\\
\theta(0)=&\pi/2\\
\phi(0)=&0\\
\end{array}
\right.
\qquad
\left \{
\begin{array}{cc}
\Delta_{yy}(\bq,0)=&V\\
\Delta_{zz}(\bq,0)=&V\\
\Delta_{ab}(\bq,0)=&0\\
\end{array}
\right.
\ee


\end{document}